\documentclass[pra,letterpaper,twocolumn,showpacs,preprintnumbers,nofootinbib,amsmath,amssymb,superscriptaddress]{revtex4}

\usepackage{amsmath,epsfig}
\usepackage{amsthm,amsfonts}

\newcommand{\ketsub}[2]{\vert#1\rangle_{#2}}
\newcommand{\brasub}[2]{\!\,_{#2}\!\langle#1\vert}
\newcommand{\kett}[1]{\vert#1\rangle\!\rangle}
\newcommand{\braa}[1]{\langle\!\langle#1\vert}
\newcommand{\braakett}[2]{\langle\!\langle#1\vert#2\rangle\!\rangle}
\newcommand{\tr}{\textrm{tr}}
\newcommand{\E}{\mathcal{E}}
\newcommand{\R}{\mathcal{R}}
\newcommand{\LL}{\mathcal{L}}
\newcommand{\HH}{\mathcal{H}}
\newcommand{\KK}{\mathcal{K}}
\newcommand{\SSS}{\mathcal{S}}
\newcommand{\AAA}{\mathcal{A}}

\newcommand{\CC}{C_{E,\E}}
\newcommand{\Ol}[1]{\overline{#1}}

\newcommand{\Yi}{Y^{(0)}}
\newcommand{\Zi}{Z^{(0)}}

\newtheorem{Pauli_Channel}{Theorem}
\newtheorem{close_isom}[Pauli_Channel]{Lemma}

\newtheorem{Gersgorin}[Pauli_Channel]{Theorem}

\input{Qcircuit}

\begin{document}
\title{Structured Near-Optimal Channel-Adapted Quantum Error Correction}
\author{Andrew S. Fletcher}\email{fletcher@ll.mit.edu}
\affiliation{Laboratory for Information and Decision Systems, Massachusetts Institute of Technology, 77 Massachusetts Ave. Cambridge, MA 02139}
\affiliation{MIT Lincoln Laboratory, 244 Wood St. Lexington, MA 02420}
\author{Peter W. Shor}\email{shor@math.mit.edu}
\affiliation{Department of Mathematics, Massachusetts Institute of Technology, 77 Massachusetts Ave. Cambridge, MA 02139}

\author{Moe Z. Win}\email{moewin@mit.edu}
\affiliation{Laboratory for Information and Decision Systems, Massachusetts Institute of Technology, 77 Massachusetts Ave. Cambridge, MA 02139}


\begin{abstract}

We present a class of numerical algorithms which adapt a quantum error correction scheme to a channel model.  Given an encoding and a channel model, it was previously shown that the quantum operation that maximizes the average entanglement fidelity may be calculated by a semidefinite program (SDP), which is a convex optimization.  While optimal, this recovery operation is computationally difficult for long codes.  Furthermore, the optimal recovery operation has no structure beyond the completely positive trace preserving (CPTP) constraint.  We derive methods to generate structured channel-adapted error recovery operations.  Specifically, each recovery operation begins with a projective error syndrome measurement.  The algorithms to compute the structured recovery operations are more scalable than the SDP and yield recovery operations with an intuitive physical form.  Using Lagrange duality, we derive performance bounds to certify near-optimality.

\end{abstract}
\maketitle

\section{Introduction}

All physical implementations of quantum information processing systems must incorporate a scheme to mitigate the effects of noise.  The most common method for quantum error correction (QEC) is analogous to classical digital error correction schemes.  The system of interest is encoded into a subspace of a larger quantum system by means of a quantum code.  After passing through a noisy channel, a syndrome measurement projects errors onto orthogonal subspaces from which the original quantum state can be recovered.  The first quantum error correcting codes demonstrated that such methods could correct arbitrary single qubit errors\cite{Sho:95,Ste:96a,CalSho:96}.  These generic methods enabled a whole range of study in quantum error correction, particularly as it applies to fault-tolerant quantum computing.

The generic approach has its drawbacks, however.  Most notably, quantum codes impose a severe amount of overhead to correct for arbitrary errors.  As an example, the shortest block code that corrects an arbitrary qubit error embeds one qubit into five\cite{BenDivSmoWoo:96,LafMiqPazZur:96}.  As scaling to many qubits is one of the principal barriers to building a working quantum computer, any efforts to improve the efficiency of error recovery are of great interest.

Several recent efforts have explored an optimization-based approach to quantum error recovery\cite{FleShoWin:07,KosLid:06,ReiWer:05,YamHarTsu:05}.  In each case, rather than correcting for arbitrary single qubit errors, the error recovery scheme was adapted to a model for the noise, with the goal to maximize the fidelity of the operation.  In \cite{FleShoWin:07}, a semidefinite program (SDP) was used to maximize the entanglement fidelity, given a fixed encoding and channel model.  In \cite{KosLid:06} and \cite{ReiWer:05}, encodings and decodings were iteratively improved using the performance criteria of ensemble average fidelity and entanglement fidelity, respectively.  A sub-optimal method for minimum fidelity, using an SDP, was proposed in \cite{YamHarTsu:05}.  An analytical approach to channel-adapted recovery based on the pretty-good measurement and the average entanglement fidelity was derived in \cite{BarKni:02}.  The main point of each scheme was to improve error corrective procedures by adapting to the physical noise process.

As in \cite{FleShoWin:07}, we choose to focus our channel-adapted efforts on the recovery operation.  While channel-adaptation can be advantageous in both the encoding and the recovery operations, the optimization problem has a significantly nicer form when one of the two is held fixed.  The numerical tools we develop can be used for either half the problem; focusing on quantum error recovery (QER) operations illustrates nearly all of the important numerical procedures.

The optimization approach to quantum error recovery demonstrates the utility of channel-adaptivity.  Such efforts have shown that quantum error correction designed for generic errors can be inefficient in the face of a particular noise process.  Since overhead in physical quantum computing devices is challenging, it is advantageous to maximize error recovery efficiency.

Recovery operations generated through convex optimization methods suffer two significant drawbacks.  First, the dimensions of the optimization problem grow exponentially with the length of the code, limiting the technique to short codes.  Secondly, the optimal operation, while physically legitimate, may be quite difficult to implement.  The optimization routine is constrained to the set of completely positive, trace preserving (CPTP) operations, but is not restricted to more easily implemented operations.

In this paper, we describe efforts to determine near-optimal channel-adapted quantum error recovery procedures that overcome the drawbacks of optimal recovery.  We impose an intuitively satisfying structure on the recovery operation and seek to optimize performance.  While still numerical procedures, the result is a class of algorithms that is less computationally intensive than the SDP and which yields recovery operations of an intuitive and potentially realizable form.

\section{Channel-adapted Recovery}

To adapt quantum error recovery to a specific channel model, we must first determine a measure of performance.  As detailed in \cite{FleShoWin:07,KosLid:06}, both entanglement fidelity ($F_e$) and ensemble average fidelity ($\bar{F}$) yield convex optimization problems.\footnote{Ensemble average fidelity yields a convex optimization problem if and only if the states in the ensemble are pure.}   As both measures may be of interest, we will use the \emph{average entanglement fidelity} of \cite{BarKni:02}, from which either of the above are a special case.  Average entanglement fidelity is defined for a channel $\AAA$ and an ensemble $E$ of states $\{\rho_i\}$ with prior probabilities $p_i$ as
\begin{equation}
  \bar{F}_e(E,\AAA)=\sum_ip_iF_e(\rho_i,\AAA)=\sum_{i,k}p_i|\tr(\rho_iA_k)|^2,
\end{equation}
where $\{A_k\}$ are the Kraus elements for the CPTP map $\AAA$.

\begin{figure}[bth]
  \begin{center}
    \includegraphics[width=\hsize]{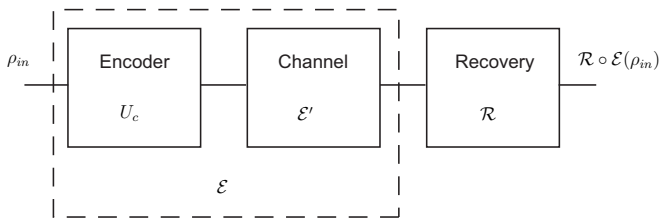}
  \end{center}
  \caption{Quantum error correction block diagram.  For channel-adapted recovery, the encoding isometry $U_C$ and the channel $\E'$ are considered as a fixed operation $\E$ and the recovery $\R$ is chosen according to the design criteria.}\label{fig:block diagram}
\end{figure}

Consider the simple block diagram for a QEC system given in Fig.~\ref{fig:block diagram}.  We begin with a fixed model, labeled $\E'$, to describe the physical noise process.  In designing a QEC procedure, we can choose the encoding $U_C$ and the recovery operation $\R$.  By holding either the encoding or the recovery operation as fixed, optimizing the other can be cast as a convex optimization problem \cite{KosLid:06,ReiWer:05,FleShoWin:07}.    As done in \cite{KosLid:06,ReiWer:05}, one can iteratively optimize a encoding/recovery scheme.  In this paper, we focus our attentions on the efficacy of adapting the recovery operation and defer iterative optimization to subsequent work.  We illustrate our emphasis on the recovery block by considering both the encoding $U_C$ and the channel $\E'$ as a combined operation $\E$.

It is useful to note the dimensions of the various operations in Fig.~\ref{fig:block diagram}.  We define two Hilbert spaces $\HH_S$ and $\HH_C$, which refer to source and the code spaces, respectively.  These have dimensions $d_S$ and $d_C$.  The combined encoding and channel $\E$ therefore maps density matrices in $\LL(\HH_S)$ to $\LL(\HH_C)$, where $\LL(\HH)$ refers to the space of bounded linear operators on $\HH$.  Our use of the fidelity implies that $\R$ maps from $\LL(\HH_C)$ to $\LL(\HH_S)$, \emph{i.e.}~$\R$ performs a decoding.  This is mostly for computational convenience as $d_S<d_C$.

Channel-adapted recovery selects an operation $\R$ to maximize $\bar{F}_e(E,\R\circ\E)$.  As shown in \cite{FleShoWin:07}, exact maximization can be accomplished via the convex optimization routine of semidefinite programming (SDP).  For the remainder of the paper, we will discuss routines to approach the optimum channel-adapted recovery through a more computationally feasible method.  In several cases, the routines also yield an intuitive form for the recovery operation.

We will make use of a convenient isomorphism in which bounded linear operators are represented by vectors and denoted with the symbol $\kett{\cdot}$.  While there are several choices for this isomorphism\cite{DarLop:01,Hav:03}, including most intuitively a ``stacking'' operation, we will follow the conventions of \cite{Tys:03} (also \cite{YamHarTsu:05}) which result in an isomorphism that is independent of the choice of basis.  For convenience, we will restate the relevant results here.

Let $A=\sum_{ij}a_{ij}\ket{i}\bra{j}$ be a bounded linear operator from $\HH$ to $\KK$ (\emph{i.e.}~$A\in\mathcal{L}(\KK,\HH))$, where $\{\ket{i}\}$ and $\{\ket{j}\}$ are bases for $\KK$ and $\HH$, respectively.  Let $\HH^*$ be the dual of $\HH$. This is also a Hilbert space, generally understood as the space of \emph{bras} $\bra{j}$.  If we relabel the elements as $\overline{\ket{j}}=\bra{j}$, then we represent $A$ as a vector in the space $\KK\otimes\HH^*$ as
\begin{equation}\label{eq:basis-free doubleket}
  \kett{A}=\sum_{ij}a_{ij}\ket{i}\overline{\ket{j}}.
\end{equation}

It is useful to note the following facts.  The inner product $\braakett{A}{B}$ is the Hilbert-Schmidt inner product $\tr A^\dagger B$.  Also, the partial trace $\tr_{\KK}\kett{A}\braa{B}=\Ol{AB^\dagger}$.  Finally, index manipulation yields the relation $A\otimes \overline{B}\kett{C}=\kett{ACB^\dagger}$, where $\overline{B}$ is the conjugate of $B$ such that $\overline{B\ket{\psi}}=\overline{B}\hspace{2pt}\overline{\ket{\psi}}$ for all $\ket{\psi}$.

These relations lead directly to a convenient representation of a CPTP operation $\AAA:\mathcal{L}(\HH)\mapsto\mathcal{L}(\KK)$ in terms of a positive semidefinite (p.s.d.) operator $X_\AAA\in\mathcal{L}(\KK\otimes\HH^*)$\cite{Cho:75,DarLop:01,Hav:03,Cav:99,Dep:67}.  The p.s.d.~operator is calculated from the Kraus elements $\{A_k\}$ of $\AAA$ as
\begin{equation}\label{eq:Choi matrix}
  X_\AAA=\sum_k\kett{A_k}\braa{A_k}.
\end{equation}
We will refer to $X_\AAA$ as the Choi matrix for $\AAA$, although most derivations do not use the basis-free free double-ket of (\ref{eq:basis-free doubleket}).  The operation output is given by $\AAA(\rho)=\tr_{\HH^*} I\otimes\overline{\rho} X_\AAA$ and the CPTP constraint requires that $X_\AAA\geq 0$ and $\tr_\KK X_A=I$.

In terms of the Choi matrix, the average entanglement fidelity can be written as $\bar{F}_e(E,\AAA)=\sum_i p_i \braa{\rho_i}X_\AAA\kett{\rho_i}$.  From this expression, we can derive the dependence of $\bar{F}_e(E,\R\circ\E)$ on $\R$ as
\begin{eqnarray}\nonumber
  \bar{F}_e(E,\R\circ\E)&=& \sum_i p_i \braa{\rho_i}X_{\R\circ\E}\kett{\rho_i}\\\nonumber
  &=& \sum_i p_i \tr X_\R (\sum_k \kett{\rho E_k^\dagger}\braa{\rho E_k^\dagger})\\
  &=& \tr X_\R \CC,\label{eq:fidelity max}
\end{eqnarray}
where $\CC=\sum_{ik} p_i \kett{\rho E_k^\dagger}\braa{\rho E_k^\dagger}$ encapsulates both the input ensemble $E$ and the channel (with encoding) $\E$.  It was shown in \cite{FleShoWin:07,KosLid:06} the the optimum $X_\R$ satisfying the CPTP constraint can be calculated via semidefinite programming.

\section{EigQER algorithm}\label{sec:EigQER}

To achieve a near-optimal QER operation, an algorithm must have a methodology to approach optimality while still satisfying the CPTP constraints.  Furthermore, to ease implementation of such a recovery, we can impose structure to maintain relative simplicity.

Let us begin by considering the structure of a standard QEC recovery operation.  QEC begins by defining a set of correctable errors, \emph{i.e.}~errors that satisfy the quantum error correction conditions.  To correct this set, we construct the recovery operation by defining a projective syndrome measurement.  Based on the detected syndrome, the appropriate unitary rotation restores the information to the code space, thereby correcting the error.  This intuitive structure -- projective measurement followed by unitary syndrome recovery -- provides a simple geometric picture of error correction.  Furthermore, it is a  relatively straightforward task to  translate such a recovery operation into a quantum circuit representation.

Let us impose the same constraint on the channel-adapted recovery operation.  We construct an operation with operator elements that are a projective syndrome measurement followed by a classically controlled unitary operation.  Thus the operator elements can be written $\{R_k=U_kP_k\}$ where $P_k$ is a projection operator.  While we could merely constrain $U_k$ to be unitary, we will instead continue with the convention that the recovery operation performs a decoding: $\R:\LL(\HH_C)\mapsto\LL(\HH_S).$  Under this convention, $U_k\in\LL(\HH_C,\HH_S)$ and $U_k^\dagger U_k=I$.  In words, both $U_k^\dagger$ and $R_k^\dagger$ are isometries.

The CPTP constraint
\begin{eqnarray}
  I&=&\sum_kR_k^\dagger R_k\\
  &=& \sum_k P_kU_k^\dagger U_k P_k\\
  &=& \sum_k P_k\label{eq:EigQER CPTP}
\end{eqnarray}
is satisfied if and only if the projectors span $\HH_C$.  To satisfy the CPTP constraint, therefore,  $\{P_k\}$ must partition $\HH_C$ into orthogonal subspaces, each identified with a correction isometry\footnote{In fact, $U_k^\dagger$ is the isometry.  For ease of explication, we will refer to $U_k$ as an isometry as well.} $U_k$.

Since the $\{P_k\}$ project onto orthogonal subspaces, we see that $R_j^\dagger R_k=\delta_{jk}P_k$.  From this we conclude that $\{\kett{R_k}\}$ are an orthogonal set and thus are eigenvectors of the Choi matrix $X_\R$.  The  eigenvalue $\lambda_k$ associated with $\kett{R_k}$ is the rank of $P_k$ and is thus constrained to be an integer.  Furthermore, since $U_k$ restores the $k^{th}$ syndrome to $\HH_S$, $\lambda_k \leq d_S$.

We can conceive of a `greedy' algorithm to construct a recovery operation $\R$.  The average entanglement fidelity can be decomposed into the contributions of each individual operator element as $\braa{R_k}C_{E,\E}\kett{R_k}$.  We can construct $\R$ by successively choosing the syndrome subspace to maximize the fidelity contribution.  As long as each syndrome is orthogonal to the previously selected subspaces, the resulting operation will be CPTP and will satisfy our additional constraints.  In fact, this greediest algorithm has no immediate method for computation; the selection of the syndrome subspace to maximize the fidelity contribution has no simple form.  We propose instead a greedy algorithm to approximate this procedure.

We motivate our proposed algorithm in terms of eigen analysis.  Let us assume for the moment that the rank of each syndrome subspace is exactly $d_S$ which is the case for QEC recoveries for stabilizer codes.  By such an assumption, we know that there will be $d_C/d_S$ recovery operator elements.  Consider now the average entanglement fidelity, in terms of the eigenvectors of $X_\R$:
\begin{equation}
  \bar{F}(E,\R\circ\E)=\sum_{k=1}^{d_C/d_S}\braa{R_k}C_{E,\E}\kett{R_k}.
\end{equation}
If we were to maximize the above expression with the only constraint being a fixed number of orthonormal vectors $\kett{R_k}$, the solution would be the eigenvectors associated with the $d_C/d_S$ largest eigenvalues of $C_{E,\E}$.  In fact, the actual constraint differs slightly from this simplification, as we further must constrain $R_k^\dagger$ to be an isometry (\emph{i.e.} $R_kR_k^\dagger=I$).  The analogy to eigen-analysis, however, suggests a computational algorithm which we dub `EigQER' (for eigen quantum error recovery).  We use the eigenvectors of $\CC$ to determine a syndrome subspace with a large fidelity contribution.

The algorithm proceeds as follows:
\begin{enumerate}
\item {Initialize $C_1=C_{E,\E}$.}\\For the $k^{th}$ iteration:
  \item Determine $\kett{X_k}$, the eigenvector associated with the largest eigenvalue of $C_k$.
    \item Calculate $R_k^\dagger$, the isometry `closest' to $X_k^\dagger$ via the singular value decomposition.  Call $R_k$ an operator element of $\R$.
    \item Determine $C_{k+1}$ by projecting out of $C_k$ the support of $R_k$.
    \item Return to step 2 until the recovery operation is complete.
\end{enumerate}
The EigQER algorithm is guaranteed to generate a CPTP recovery operation, and will satisfy the criterion that it can be implemented by a projective syndrome measurement followed by a syndrome dependent unitary operation.

Steps 2 and 3 in the above algorithm require further exposition.  Given an operator $X\in\LL(\HH_C,\HH_S)$, what is the closest isometry $R_k$?  A straightforward answer uses the norm derived from the Hilbert-Schmidt inner product where $\|A\|^2=\tr A^\dagger A$.  We will now allow the rank of $k^{th}$ subspace to be $d_k\leq d_S$.\footnote{Inclusion of reduced rank subspaces may seem unnecessary or even undesirable - after all, such a projection would collapse superpositions within the encoded information.  We allow the possibility since such operator elements are observed in the optimal recovery operations of \cite{FleShoWin:07}.}  Thus $R_kR_k^\dagger=I_{d_k}$ where $I_{d_k}$ is a diagonal operator with the 1 as the first $d_k$ diagonal matrix elements and 0 for the rest.
We have the minimization problem
\begin{eqnarray}\label{eq:closest isometry}
\min_{R_k} \tr(X-R_k)^\dagger(X-R_k) \textrm{ such that } R_kR_k^\dagger=I_{d_k}.
\end{eqnarray}

We will state the solution as the following lemma.
\begin{close_isom}\label{thm:closest isometry}
  Let $X$ be an operator with singular value decomposition $X=U\Sigma V^\dagger$.  The rank $d$ isometry $R$ that minimizes the Hilbert-Schmidt norm difference $\|X-R\|$ is given by $R=UI_{d_I}V^\dagger$.
\end{close_isom}
\begin{proof}
Let $\mathcal{U}_d$ be the set of rank $d$ isometries;  that is $\mathcal{U}_d=\{U|U^\dagger U =I_d\}$.
We wish to find the $R^\dagger \in\mathcal{U}$ that minimizes $\tr(X-R)^\dagger(X-R)$.  Since  this can be written as
\begin{equation}
\tr(X-R)^\dagger(X-R)=\tr X^\dagger X + \tr R^\dagger R - \tr(X^\dagger R+R^\dagger X)
\end{equation}
and $\tr R^\dagger R=d$, an equivalent problem is
\begin{eqnarray}\label{eq:closest isometry2}
\max_{R\in\mathcal{U}} \tr(X^\dagger R+R^\dagger X)=\max_{R\in\mathcal{U}}\tr(V\Sigma U^\dagger R+R^\dagger U\Sigma V^\dagger),
\end{eqnarray}
where we have replaced $X$ with its singular value decomposition.

We can simplify the above expression by noting that $C^\dagger=U^\dagger R\in\mathcal{U}$.  We can thus equivalently maximize the following expression over $C^\dagger\in\mathcal{U}$:
\begin{eqnarray}
\tr(V\Sigma C^\dagger+C\Sigma V^\dagger)&=&\tr\Sigma (C^\dagger V+V^\dagger C)\\
\label{eq:closest_isom_proof2}
&=& \sum_{i=1}^d \sigma_i (c_i^\dagger v_i+v_i^\dagger c_i)\\
&=& 2\sum_{i=1}^d \sigma_i \textrm{Re}\{v_i^\dagger c_i\}\\
&\leq& 2\sum_{i=1}^d \sigma_i |v_i^\dagger c_i|\\
&\leq& 2\sum_{i=1}^d \sigma_i \|v_i\|\|c_i\|\\
&=& 2\sum_{i=1}^d \sigma_i.
\end{eqnarray}
In (\ref{eq:closest_isom_proof2}), $\sigma_i$ is the $i^{th}$ largest singular value of $X$ and $v_i$ and $c_i$ are the $i^{th}$ columns of $V$ and $C$, respectively.
We have used the fact that $\Sigma$ is a diagonal matrix of the singular values in descending order.  The inequality is saturated when $c_i=v_i$, which also implies that $C= V I_d\Rightarrow R=UI_dV^\dagger$.
\end{proof}

One item not mentioned above is the determination of the desired rank $d_k$.  In our implementation of EigQER, this is accomplished by setting a relatively high threshold on the singular values of $X$.  We only considered singular values such that $\sigma^2\geq.05$.  This \emph{ad hoc} value was chosen as it led to acceptable numerical results in the examples.

We turn now to step 3 of the EigQER algorithm.  Recall that the CPTP constraint as written in (\ref{eq:EigQER CPTP}) requires that the syndrome subspaces are mutually orthogonal.  Thus, the syndrome measurement for the $k^{th}$ iteration must be orthogonal to the first $k-1$ iterations: $P_{k}P_i=0$ for $i<k$.  We satisfy this constraint by updating the data matrix $C_{k-1}$.

To understand the update to $C_{k-1}$, recall that the first step of the $k^{th}$ iteration is the computation of the dominant eigenvector $\kett{X_k}$.  To satisfy the constraint, we require that
\begin{equation}
X_kP_i=0\Leftrightarrow\kett{X_kP_i}=I\otimes\overline{P_i}\kett{X_k}=0
\end{equation}
for $i<k$.  All $\kett{X}$ for which this is not satisfied should be in the nullspace of $C_k$.  Thus, after each iteration we update the data matrix as
\begin{equation}
  C_{k}=(I-I\otimes \overline{P_{k-1}})C_{k-1}(I-I\otimes \overline{P_{k-1}}).
\end{equation}

The algorithm terminates when the recovery operation is complete, \emph{i.e.} $\sum_k R_k^\dagger R_k=\sum_kP_k=I$.  Given the structure of the recovery operations, this can be determined with a simple counter that is increased by $d_k$ at each step $k$.  When the counter reaches $d_C$, the recovery is complete.

\begin{figure}
  \begin{center}
  \includegraphics[width=\columnwidth]{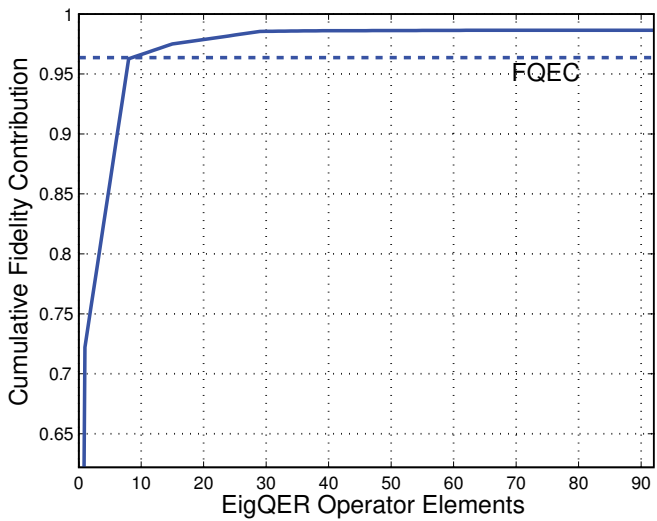}
    \caption[Fidelity contribution of EigQER recovery operators for the amplitude damping channel and the Steane code.]{Fidelity contribution of EigQER recovery operators for the amplitude damping channel ($\gamma=.09$) and the Steane code.  Notice that the QEC performance is equaled with only 8 operator elements, and the relative benefit of additional operators goes nearly to zero after 30.}
    \label{fig:EigQER greedy}
  \end{center}
\end{figure}

In fact, the greedy nature of EigQER allows early termination of the above algorithm.  Each $R_k$ contributes $\braa{R_k}C_{E,\E}\kett{R_k}$ to the average entanglement fidelity.  Since the algorithm seeks to maximize its gain at each step, the performance return of each $R_k$ diminishes as $k$ grows.  This is illustrated in Fig.~\ref{fig:EigQER greedy}, where we show the cumulative contribution for each recovery operator element with the Steane code and the amplitude damping channel.  The greedy construction results in simplifications in both computation and implementation.  When the contribution $\braa{R_k}\CC\kett{R_k}$ passes below some selected threshold, the algorithm may terminate and thus reduce the computational burden.  This results in an under-complete recovery operation where $\sum_k R_k^\dagger R_k\leq I$.  An under-complete specification for the recovery operation may significantly reduce the difficulty in physically implementing the recovery operation.  In essence, an under-complete recovery operation will have syndrome subspaces whose occurrence is sufficiently rare that the recovery operation may be left as a `don't care.'

Before we consider examples of EigQER recovery performance, we should say a few words about the algorithm complexity when channel adapting an $[n,k]$ code.  The SDP of \cite{FleShoWin:07,KosLid:06} to calculate the optimal recovery operation has $4^{n+k}$ complex optimization variables constrained to a semidefinite cone with a further $4^k$ equality constraints.  From \cite{BoyVan:B04}, an SDP with $n$ variables and a $p\times p$ semidefinite matrix constraint requires $\mathcal{O}(\max \{np^3,n^2p^2,n^3\})$ flops per iteration (with typically 10-100 iterations necessary).  For our case, this yields $\mathcal{O}(2^{5(n+k)})$ flops per iteration.

For the EigQER operation, the dominant computation is the calculation of $\kett{X_k}$, the eigenvector associated with the largest eigenvalue of $C_k$.  $C_k$ is a $2^{n+k}\times 2^{n+k}$ dimensional matrix, but the eigenvector has only $2^{n+k}$ dimensions.  Using the \emph{power method} for calculating the dominant eigenvector requires $\mathcal{O}(2^{2(n+k)})$ flops for each iteration of the power method.  While both problems grow exponentially with $n$, the reduced size of the eigenvector problem has a significant impact on the computational burden.

We should note that the eigenvector computation must be repeated for each operator element of $\R$.  If we were to compute all of them, not truncating early due to the diminishing returns of the greedy algorithm, this would require iterating the algorithm approximately $d_C/d_S=2^{n-k}$ times.  In fact, we have a further reduction as the algorithm iterates.  At the $j^{th}$ iteration we are calculating the dominant eigenvector of $C_j$ which lives on a $(d_C-jd_S)d_S=2^k(2^n-j2^k)$ dimensional subspace.  We can therefore reduce the size of the eigenvector problem at each iteration of EigQER.

\subsection{EigQER examples}

To demonstrate the value of the EigQER algorithm, we consider several channels and codes; we would like to consider common codes and channels with non-trivial channel-adapted recoveries.  It will be shown in the appendix that channels represented by scaled Pauli group operators yield straightforward channel-adapted recovery operations; it is therefore useful to consider non-Pauli channels.  The most common and useful such channel is the amplitude damping channel, which we will denote $\E_a$.  Amplitude damping was the example used in \cite{FleShoWin:07} to illustrate optimal QER, as well as the example for channel-adapted code design of \cite{LeuNieChuYam:97}.  The channel is a commonly encountered model, where the parameter $\gamma$ indicates the probability of decaying from state $\ket{1}$ to $\ket{0}$ (\emph{i.e.}~the probability of losing a photon).  For a single qubit, $\E_a$ has operator elements
\begin{equation}\label{eq:ampdamp}
E_0=\left [ \begin{array}{ccc} 1 & 0 \\ 0 &\sqrt{1-\gamma} \end{array} \right ]\hspace{.5 cm} \textrm{and} \hspace{.5 cm}
E_1=\left [ \begin{array}{ccc} 0 & \sqrt{\gamma} \\ 0 & 0 \end{array} \right ].
\end{equation}

The EigQER algorithm does not require a channel as simple to model as the amplitude damping channel; the optimization routine is general to any channel.  To illustrate, we consider a qubit channel that is less familiar, though with a straightforward geometric description.  We will call this the `pure states rotation' channel and label it as $\E_{ps}$.  To describe the channel, we define a pure state by its angle in the $xz$-plane: $\ket{\theta}=\cos\theta\ket{0}+\sin\theta\ket{1}$.  The channel mapping is defined by its action on two pure states an angle $\theta$ apart, symmetric about the $z$-axis.  When $\ket{\pm\theta/2}$ is input to the channel, the result is $\ket{\pm(\theta-\phi)/2}$, also as a pure state.  Thus, these two states are rotated toward each other by $\phi$.  Any other state input to the channel will emerge mixed.  The operator elements for this channel can be written as
\begin{eqnarray}\label{eq:purestates channel}
\E_{ps}&\sim&\left\{
\alpha \begin{bmatrix}
  \cos\frac{\theta-\phi}{2}\sin\frac{\theta}{2} & \pm\cos\frac{\theta-\phi}{2}\cos\frac{\theta}{2}\\
  \pm\sin\frac{\theta-\phi}{2}\sin\frac{\theta}{2} &   \sin\frac{\theta-\phi}{2}\cos\frac{\theta}{2}
\end{bmatrix},
\beta \begin{bmatrix}
  \frac{\cos\frac{\theta-\phi}{2}}{\cos\frac{\theta}{2}} & 0\\
  0 & \frac{\sin\frac{\theta-\phi}{2}}{\sin\frac{\theta}{2}}
\end{bmatrix}\right\},
\end{eqnarray}
where $\alpha$ and $\beta$ are constants chosen to satisfy the CPTP constraint.

The pure states rotation channel has multiple parameters which characterize its behavior.  $\theta$ indicates the initial separation of the targeted states.  $\phi$, the amount of rotation, clearly parameterizes the `noise strength' as $\phi=0$ indicates no decoherence while $\phi=\theta$ is strong decoherence.  Furthermore, we have chosen the target states to be symmetric about the $z$-axis, but this is only for clarity in stating the channel; any alternate symmetry axis may be defined.  Furthermore, a similar channel with asymmetric rotations $\phi_1$ and $\phi_2$ may be defined.  This, however, corresponds to a symmetric channel followed by a unitary rotation.  While less physically motivated than amplitude damping, the pure state rotation channel model provides an extended set of qubit channels which are not represented with Pauli group operator elements.  We will look at examples of this channel where $\theta= 5\pi/12$.  There is no particular significance to this choice; it merely illustrates well the principles of channel-adapted QEC.

Since the EigQER algorithm is more computationally scalable than the SDP, we can consider channel-adapted QER for several codes.  We compare the EigQER recovery performance to the optimal channel-adapted recovery performance for the 5 qubit stabilizer code\cite{BenDivSmoWoo:96,LafMiqPazZur:96}.  We also compare the EigQER performance for the 5 qubit code, the 7 qubit Steane code\cite{Ste:96a,CalSho:96}, and the 9 qubit Shor code\cite{Sho:95}.  All comparisons consider an ensemble $E$ of qubit states that are in the completely mixed state $\rho=I/2$.

\begin{figure}
\includegraphics[width=\columnwidth]{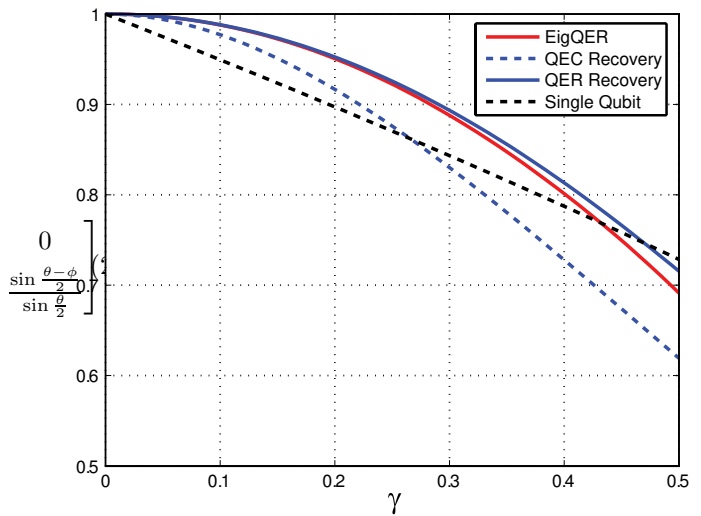}
  \caption[EigQER and Optimal QER for the amplitude damping channel and the five qubit stabilizer code.]{EigQER and Optimal QER for the amplitude damping channel and the five qubit stabilizer code.  EigQER nearly duplicates the optimal channel-adapted performance, especially for lower noise channels (small $\gamma$).}
\label{fig:AmpDamp5_eigQER}
\end{figure}

Figure \ref{fig:AmpDamp5_eigQER} compares the performance of the EigQER algorithm to the optimal QER recovery for the case of the five qubit stabilizer code and the amplitude damping channel.  Also included are the generic QEC recovery and the entanglement fidelity of a single qubit acted upon by $\E_a$ (\emph{i.e.}~no error correction performed).  From this example we observe that the EigQER performance nearly achieves the optimum, especially for the values of $\gamma$ below $.4$.  For higher $\gamma$, the EigQER performance begins to diverge, but this is less important as that region is one in which even the optimal QER lies below the fidelity of a single qubit obtainable with no error correction.

\begin{figure}
\includegraphics[width=\columnwidth]{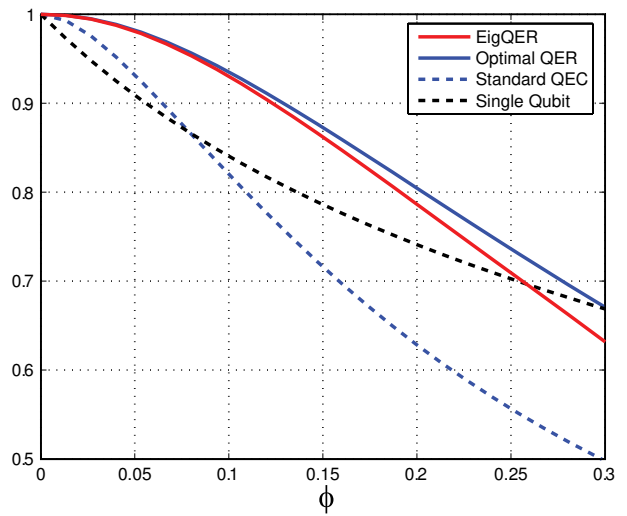}
  \caption[EigQER and Optimal QER for the pure state rotation channel with $\theta=5\pi/12$ and the five qubit stabilizer code.]{EigQER and Optimal QER for the pure state rotation channel with $\theta=5\pi/12$ and the five qubit stabilizer code.  EigQER nearly duplicates the optimal channel-adapted performance, especially for lower noise channels (small $\phi$).}
  \label{fig:PureStates5_eigQER}
\end{figure}

Figure \ref{fig:PureStates5_eigQER} compares EigQER and optimal QER for the five qubit stabilizer code and the pure state rotation channel with $\theta=5\pi/12$.  We see again that the EigQER algorithm achieves a recovery performance nearly equivalent to the optimum, especially as the noise level approaches $0$.

\begin{figure}
\includegraphics[width=\columnwidth]{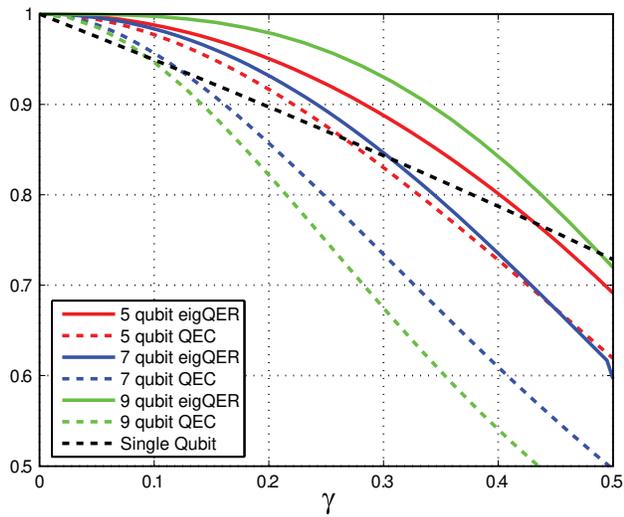}
  \caption[EigQER and standard QEC recovery performance for the five, seven, and nine qubit codes and the amplitude damping channel.]{EigQER and standard QEC recovery performance for the five, seven, and nine qubit codes and the amplitude damping channel.  Note that generic QEC performance decreases for longer codes, as multiple qubit errors become more likely.  While the EigQER performance for the nine qubit Shor code is excellent, the seven qubit Steane code shows only modest improvement, with performance similar to the generic five qubit QEC recovery.}
\label{fig:AmpDamp579_eigQER}
\end{figure}

Figure \ref{fig:AmpDamp579_eigQER} demonstrates the performance of several codes and the amplitude damping channel.  We compare the EigQER performance for the five, seven, and nine qubit codes, contrasting each with the generic QEC performance.  Notice first the pattern with the standard QEC recovery: the entanglement fidelity decreases with the length of the code.  The five qubit stabilizer code, the Steane code, and the Shor code are all designed to correct a single error on an arbitrary qubit, and fail only if multiple qubits are corrupted.  For a fixed $\gamma$, the probability of a multiple qubit error rises as the number of physical qubits $n$ increases.

The QEC performance degradation with code length is a further illustration of the value of channel adaptivity.  All three codes in Figure \ref{fig:AmpDamp579_eigQER} contain one qubit of information, so longer codes include more redundant qubits.  Intuitively, this should better protect the source from error.  When we channel adapt, this intuition is confirmed for the Shor code, but not for the Steane code. In fact, the EigQER entanglement fidelity for the Steane code is only slightly higher than the generic QEC recovery for the five qubit code.  From this example, it appears that the Steane code is not particularly well suited for adapting to amplitude damping errors.  We see that the choice of encoding significantly impacts channel-adapted recovery.

\begin{figure}
\includegraphics[width=\columnwidth]{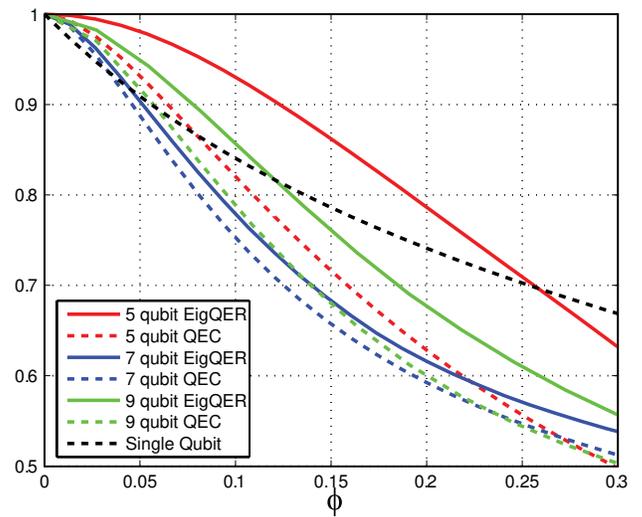}
  \caption[EigQER and standard QEC recovery performance for the five, seven, and nine qubit codes and the pure state rotation channel with $\theta=5\pi/12$.]{EigQER and standard QEC recovery performance for the five, seven, and nine qubit codes and the pure state rotation channel with $\theta=5\pi/12$.  Despite the least redundancy, the five qubit code has the best channel-adapted performance.  The Steane code appears particularly poor for this channel: both the generic QEC and the adapted recovery have lower fidelity than the other codes.}
\label{fig:PureStates579_eigQER}
\end{figure}

The effect is even more dramatically (and puzzlingly) illustrated in the pure state rotation channel.  Figure \ref{fig:PureStates579_eigQER} compares the EigQER recoveries for the five qubit, Steane, and Shor codes with $\theta=5\pi/12$.  It is interesting to see that the five qubit code outperforms each of the others despite less redundancy to protect the information.  Furthermore, both the standard QEC and channel-adapted recoveries for the Steane code perform worse than the generic recovery of the Shor code!  This suggests that the five qubit code is particularly well suited to adapt to errors of this type, while the Steane code is particularly ill-suited.  (We suspect that the Shor code with QEC recovery outperforms the Steane due to its degenerate structure.)

\section{Block SDP QER}\label{sec:BlockSDP}

The recovery operation generated by the EigQER algorithm of the preceding section is one of a broader class of quantum error recoveries.  The class is characterized by an initial projective syndrome measurement, followed by a syndrome-specific recovery operation.  The projective measurement partitions $\HH_C$ and provides some knowledge about the observed noise process.

Projective syndrome measurements for quantum error correction are tricky to design.  We wish to learn as much as possible about the error while learning as little as possible about the input state, so as not to destroy quantum superposition.  The EigQER algorithm aggressively designs the syndrome measurement, as the $R_k=U_kP_k$ structure of the operator elements implies a finality about the syndrome selection.  The outcome of the syndrome measurement completely determines the correction term $U_k$.

We can conceive of a less aggressive projective measurement.  If we projected onto larger subspaces of $\HH_C$, we would learn less about the noise but perhaps have less chance of destroying the superposition of the input state.  We could consider this an intermediate syndrome measurement, a preliminary step to further error correction.  To design a recovery operation of this type, we must have a strategy to select a projective measurement.  Given the outcome $P_k$, we must further design the syndrome recovery operation $\R_k$.  


Consider the projective syndrome measurement operator $P_k$.  For the EigQER algorithm, $P_k=R_k^\dagger R_k$ always projects onto a subspace of dimension less than or equal to the source space: $\textrm{rank}(P_k)\leq d_S$.  This is an aggressive condition that arises from constraining the subsequent syndrome recovery to be a unitary operator.  We will relax this constraint and allow an arbitrary syndrome recovery $\R_k$ for the $k^{th}$ syndrome measurement.  It turns out that we can determine the optimum such recovery $\R_k^{opt}$ via semidefinite programming, just as in \cite{FleShoWin:07}.  The intermediate syndrome measurement $P_k$ reduces the dimension of the SDP, and thus the technique is still applicable to long codes where computing the global optimum recovery is impractical.

We will demonstrate how the optimum syndrome recovery $\R_k$ can be calculated via a semidefinite program.  Let $\{P_k\}_{k=1}^{K}$ be a set of projectors such that $\sum_k P_k = I\in\HH_C$ that constitute an error syndrome measurement.  Let $\SSS_k$ be the support of $P_k$ with dimension $d_k$; it is clear that $\SSS_1\oplus\SSS_2\oplus\cdots\oplus\SSS_K=\HH_C$.  Given the occurrence of syndrome $k$, we must now design a recovery operation $\R_k:\SSS_k\mapsto\HH_S$.  $\R_k$ is subject to the standard CPTP constraint on quantum operations, but only has support on $\SSS_k$.  We may calculate the recovery $\R_k$ that maximizes the average entanglement fidelity using the SDP in a structure identical to that of \cite{FleShoWin:07} while accounting for the reduced input space:
\begin{eqnarray}\label{eq:reduced SDP}
X_{\R_k}=\arg \max_X \tr X (\CC)_k,\\
\nonumber \textrm{such that }X\geq 0,\textrm{ }\tr_{\HH_S}X=I\in\SSS_k.
\end{eqnarray}
Here, $(\CC)_k=I\otimes\overline{P_k}\CC I\otimes\overline{P_k}$ is the data matrix projected into the $k^{th}$ subspace.  Notice that $X_{\R_k}$ and $(\CC)_k$ are operators on $\HH_S\otimes\SSS_k^*$.  In contrast to $\CC$, which requires $d_S^2d_C^2$ matrix elements, $(\CC)_k$ is fully specified by $d_S^2d_k^2$ matrix elements.  By partitioning $\HH_C$ into subspaces $\{\SSS_k\}$ through a careful choice of a syndrome measurement $\{P_k\}$, we may apply semidefinite programming to high dimensional channels without incurring the full computational burden of computing the optimal recovery.  In the following sections we discuss two strategies for determining the syndrome measurement.

\subsection{Block EigQER}

The first step of an iteration of EigQER computes the dominant eigenvalue and corresponding eigenvector of $\CC$.  This eigenvector corresponds to the operator that maximizes the average entanglement fidelity gain at a single step.  While such an operator may violate the CPTP constraint for the recovery operation, it serves to identify an important subspace onto which we may project.  Indeed, the good performance of the EigQER algorithm rests on the successful identification of suitable syndrome subspaces via eigen-analysis.

An intuitive extension of this concept is to use multiple eigenvectors to specify a higher-dimension subspace.  If $\{\kett{X_m}\}_{m=1}^{M}$ are the eigenvectors corresponding to the $M$ largest eigenvalues of $\CC$, then it is reasonable to define the subspace $\SSS_1$ as the union of the support of the operators $\{X_m\}$.  We define the corresponding projector $P_1$ and calculate the syndrome recovery $\R_1$ via the SDP of (\ref{eq:reduced SDP}).  As in the EigQER algorithm, we update the data matrix $C$ by projecting out the subspace $\SSS_1$, at which point we select another set of eigenvectors.  We will refer to this algorithm as BlockEigQER.

How many eigenvectors should be selected to define a block?  A simple solution is for a fixed block size, say $M$, to be processed until the recovery is complete.  For $M=1$, BlockEigQER is identical to EigQER.  For $M=d_Sd_C$, BlockEigQER computes the optimal recovery operation, as the syndrome measurement is simply the identity operator.  For values in between, one would expect to trade off performance for computational burden.  While there is no guarantee that performance will improve monotonically, we would anticipate improved performance as $M$ increases.

\begin{figure}
\begin{center}
\begin{tabular}{c}
\includegraphics[width=.7\columnwidth]{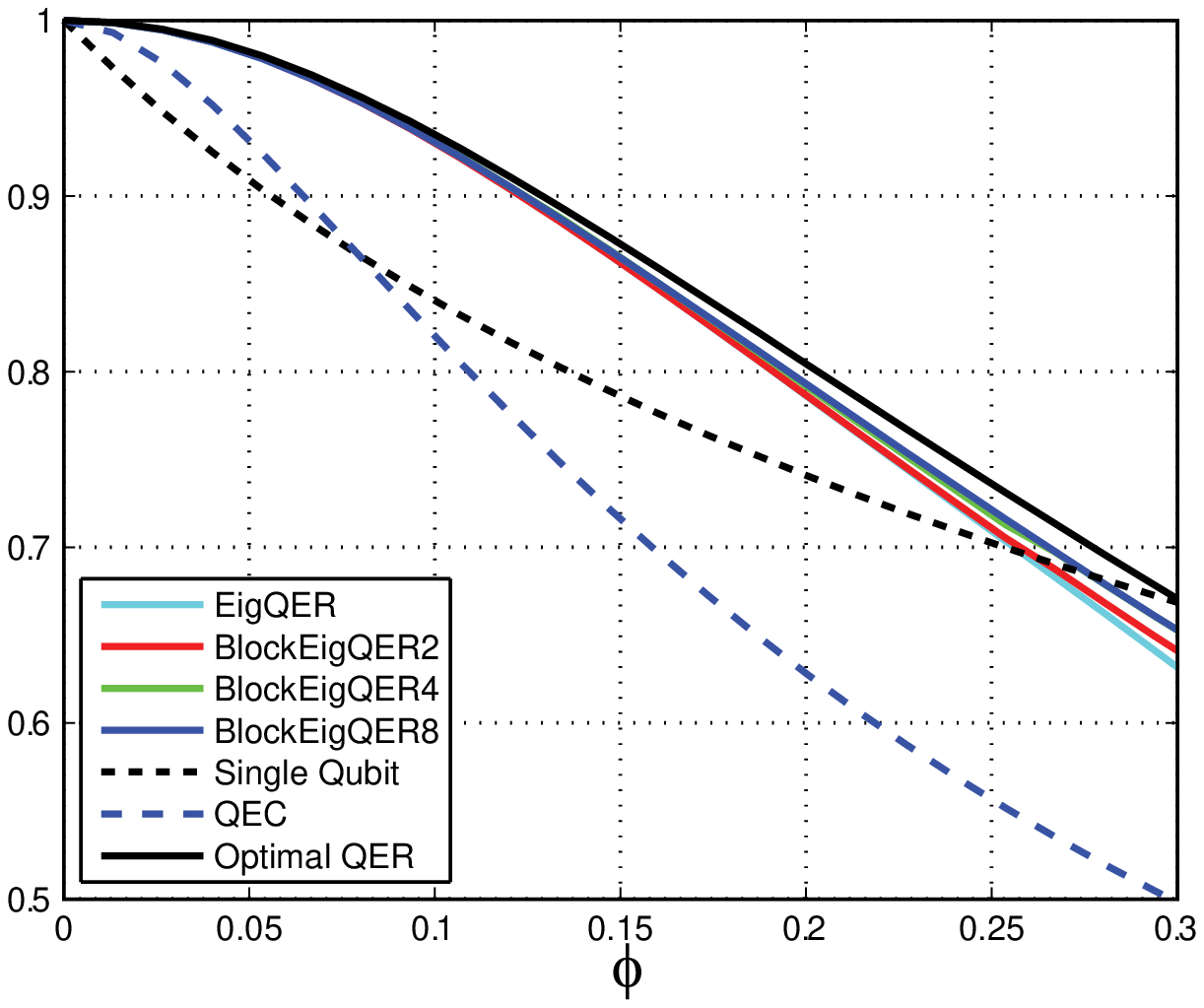}\\

(A)
\vspace{.5cm}\\
\includegraphics[width=.7\columnwidth]{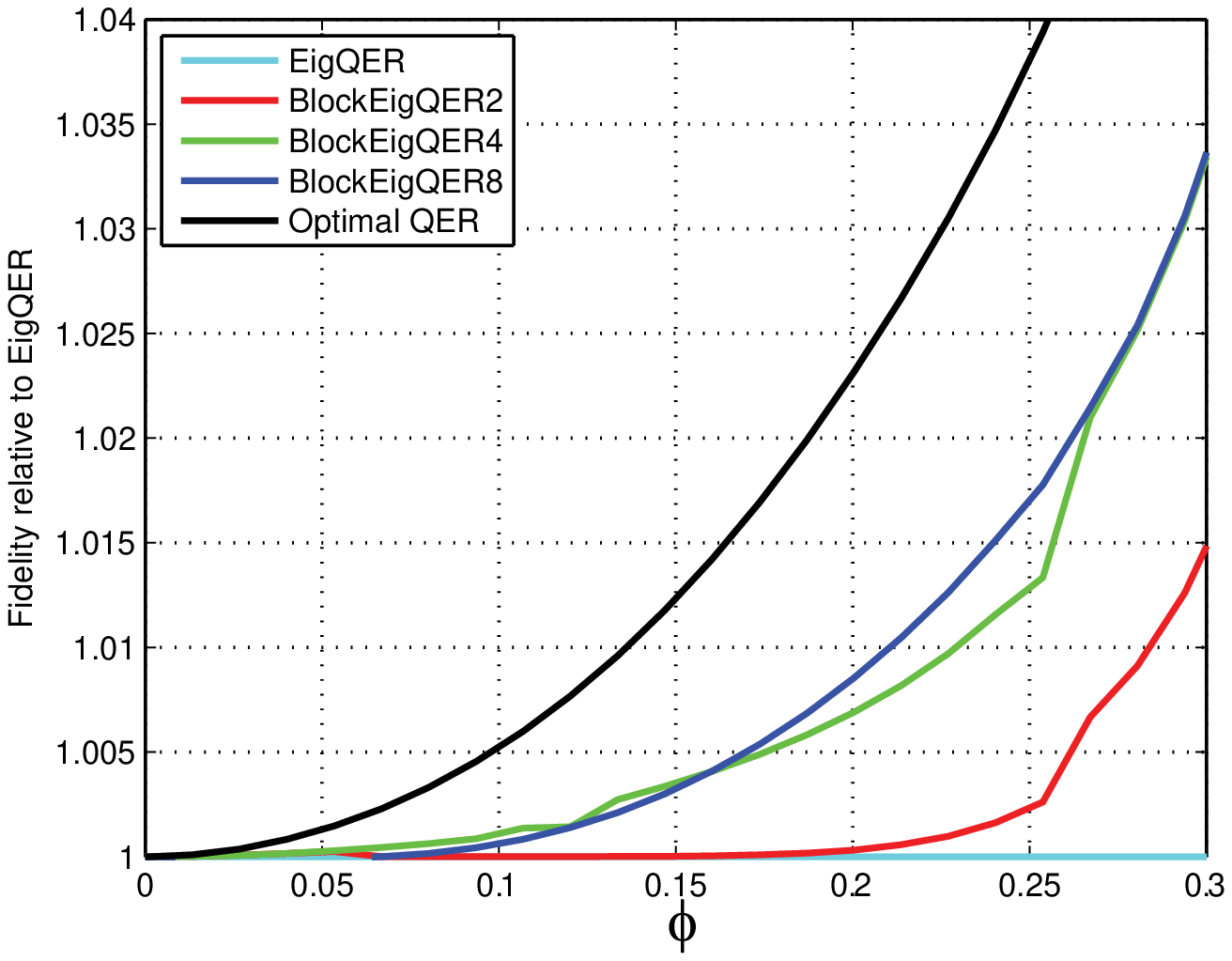}\\

(B)
\end{tabular}
\end{center}

  \caption[BlockEigQER performance for the five qubit code and the pure state rotation channel with $\theta=5\pi/12$.]{BlockEigQER performance for the five qubit code and the pure state rotation channel with $\theta=5\pi/12$.  BlockEigQER is computed with fixed block lengths of 2, 4, and 8.  In (A) we compare the entanglement fidelity to the EigQER recovery, standard QEC recovery and Single Qubit baseline.  The different block lengths have nearly indistinguishable performance from EigQER.  In (B), we compute the fidelity relative to the EigQER recovery and show that the fidelity improves by less than $4\%$ for the displayed region.  We can note, however, that longer block lengths tend to better performance.}\label{fig:BlockEigQER}
\end{figure}

We illustrate the performance for several choices of $M$ in Fig.~\ref{fig:BlockEigQER}.  We use the pure state rotation channel ($\theta=5\pi/12$) and the five qubit code with block sizes of 2, 4, and 8.  The expected improvement as $M$ increases is evident, though the gain is quite modest for noise levels of interest (below the cross-over with the single qubit recovery) and is not strictly monotonic.  The variations in performance , including the non-monotonicity, are likely the result of syndrome measurements that collapse the input superpositions.  While the eigenvectors of $\CC$ that identify the syndrome subspace generally avoid collapsing the input state, the mechanism is imperfect.

\begin{figure}
  \begin{center}
    \includegraphics[width=\columnwidth]{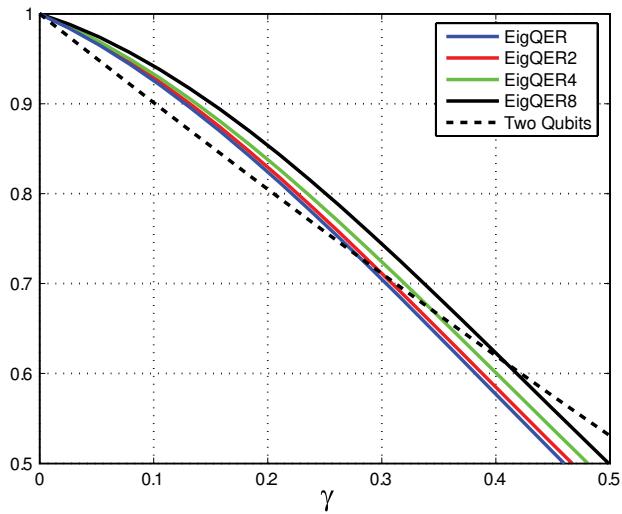}
    \caption[BlockEigQER for the amplitude damping channel and a random six qubit code.]{BlockEigQER for the amplitude damping channel and a random [6,2] code.  We compare the BlockEigQER algorithm for block sizes of 2,4, and 8 with EigQER algorithm.  We see significant performance improvement for larger block sizes, at the cost of computational and recovery complexity.  Baseline in this case is the entanglement fidelity for two qubits input to the channel without error correction.}\label{fig:Random_BlockEigQER}
  \end{center}
\end{figure}

While BlockEigQER outperforms EigQER in the $[5,1]$ code, we see in (B) of Fig.~\ref{fig:BlockEigQER} that the improvement is less than $5\%$ within the $\phi$ of interest.  We see more significant gains when we encode multiple qubits.  Consider a random $[6,2]$ encoding for the amplitude damping channel, shown in Figure \ref{fig:Random_BlockEigQER}.  In this case we see a distinct performance gain as $M$ increases and the difference is non-trivial.

Fixing the block size $M$ ignores some of the inherent symmetries in the channel and encoding.  In particular, it is quite common for $\CC$ to have degenerate eigenvalues.  By fixing the number of eigenvectors to simultaneously consider, one may inadvertently partition such a degenerate subspace according to the numerical precision of the eigen-analysis software.  To avoid this unwanted circumstance, we may select a variable block size based on the magnitude of the eigenvalues.  This approach necessitates a strategy for parsing the eigenvalues into variable size blocks which can be a tricky procedure.  Due to the modest returns of such an attempt, we have not pursued such a strategy.

While BlockEigQER shows modest performance improvements when compared to EigQER, it has one significant drawback.  Unlike EigQER, the recovery operation from BlockEigQER is not constrained to a collection of isometries.  Once the initial projective syndrome measurement is performed, the subsequent correction terms are arbitrary CPTP maps.  This may complicate attempts to physically implement such an operation.  Furthermore, BlockEigQER does not provide much more intuition for recovery design than EigQER.  For this reason, we consider BlockEigQER a numerical tool whose principal value is its incremental improvement approaching optimality.  It also prove useful for the performance bounds derived in Section \ref{sec:Dual Bounds}.

\subsection{OrderQER}

We now consider a block QER algorithm that provides intuition for error recovery design.  We are often interested in channels where each qubit is independently corrupted; thus the overall channel is the tensor product of single qubit channels.  We can use this structure to design an intuitive projective measurement.  We illustrate using the classical bit flip channel with probability of error $p$.  If a single bit of the codeword is flipped, we label this a `first order error' as the probability of such an error is $\mathcal{O}(p)$.  If two codeword bits are flipped, this is a `second order error', which occurs with probability $\mathcal{O}(p^2)$.

This intuition can easily yield a choice of syndrome subspaces $\{\SSS_k\}$.  Consider, for example, the amplitude damping channel given in (\ref{eq:ampdamp}).  Recognizing $E_1$ as the `error event,' we declare first order errors to be of the form $E^1_k=E_0\otimes\cdots E_1\otimes E_0\otimes\cdots$ where the error is on the $k^\textrm{th}$ qubit.  In this case we can declare the first order syndrome subspace to be
\begin{equation}\label{eq:first order subspace}
\SSS_1=\textrm{span}(\{\ket{E_0^{\otimes n}0_L},\ket{E_0^{\otimes n}1_L},\ket{E_1^10_L},\ket{E_1^11_L},\cdots\ket{E_n^11_L}\}),
\end{equation}
where $\ket{0_L}$ and $\ket{1_L}$ are the logical codewords for an $n$-length code.  We include the `no error' term as numerical experience suggests that the code projector $P_C$ is not always an optimal syndrome measurement.  By parallel construction, we can define the second order syndrome subspace $\SSS_2$.  While these two will probably not complete the space $\HH_C$, quite possibly we may neglect any higher orders.  Alternatively we can analyze the remaining subspace with either the SDP or the numerically simpler EigQER algorithm.  We will refer to this block SDP algorithm as OrderQER.

The SDP's for first and second order subspaces significantly reduce the dimension from the full optimal SDP, though the effect is not as dramatic as BlockEigQER.  Consider the case of the amplitude damping channel which has only two operator elements for the single qubit channel.  For an $[n,k]$ code, there is one `no error' operator and $n$ first order error operators.  This suggests that $\SSS_1$ has dimension $(n+1)d_S=(n+1)2^k$.  The SDP then has $(n+1)^22^{4k}$ optimization variables. Contrast this $n^2$ growth with the $4^n$ growth of the optimal SDP.  For second order errors, there are $\begin{pmatrix} n\\2\end{pmatrix}\approx \frac{n^2}{2}$ error operators.  The subspace $\SSS_2$ has approximate dimensions of $n^22^{k-1}$ and thus the SDP has $n^42^{4k-2}$ optimization variables.  For the $[7,1]$ Steane code, computing the full optimal SDP requires an impractical $4^7\cdot4=65536$ variables.  However, the first order SDP requires $8^22^4=1024$ variables and the actual second order SDP has $42^2\cdot4=7056$ optimization variables.  For contrast, the full SDP and the five qubit code requires 1024 optimization variables.  For the $[9,1]$ Shor code, the second order SDP has an impractical $72^2\cdot4=20736$ optimization variables.  We therefore do not use OrderQER for the Shor code.

\begin{figure}
  \begin{center}
    \includegraphics[width=\columnwidth]{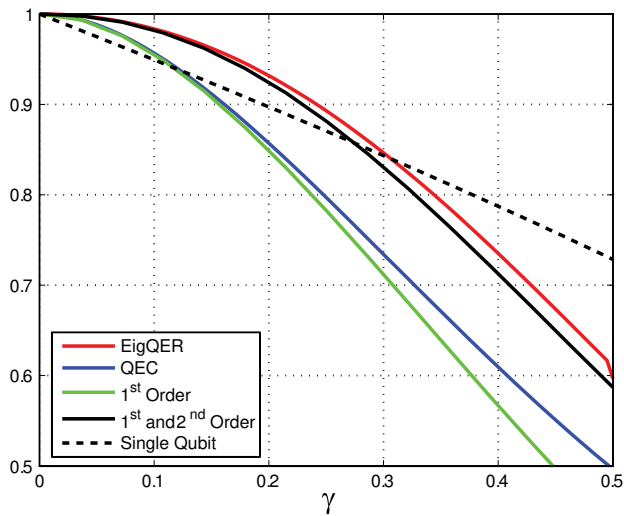}
    \caption[OrderQER recovery for the seven qubit Steane code and the amplitude damping channel.]{OrderQER recovery for the seven qubit Steane code and the amplitude damping channel.  We compare the recovery fidelity of the $1^{st}$ order error to the standard QEC performance.  The performance of the $1^{st}$ and $2^{nd}$ order recoveries together are comparable to the EigQER recovery, especially as $\gamma$ approaches 0.}\label{fig:OrderQER}
  \end{center}
\end{figure}

While the scaling of OrderQER grows quickly with $n$ making its use challenging for codes as long as nine qubits, OrderQER results provide significant insight into the mechanism of channel-adaptation.  Consider the $1^{st}$ and $2^{nd}$ order recovery performance for the Steane code and the amplitude damping channel from Figure \ref{fig:OrderQER}.  We note that the fidelity performance for the recovery from $\SSS_1$ is comparable to the performance of standard QEC, especially as $\gamma$ approaches 0.  This matches the intuition that standard QEC is correcting single qubit errors which are almost completely restricted to $\SSS_1$.  For small $\gamma$, the most likely syndrome measurement will be a Pauli $X$ or $Y$, as these characterize single qubit dampings.  These same errors are corrected by $1^{st}$ order OrderQER.   As $\gamma$ grows, the distortion from the `no error' term $E_0\otimes\cdots\otimes E_0$ becomes more pronounced and the QEC outperforms $1^{st}$ order OrderQER.

We see that $1^{st}$ and $2^{nd}$ order recovery performance is quite comparable to the EigQER performance.  Thus, the performance gains observed for channel adapted QER can be understood as corrections of higher order errors.  Since $\SSS_1$ has dimension significantly less than $d_C$ and yet approximates the QEC recovery performance, it is only reasonable that the remaining redundancy of the code can be exploited to protect from further error.

\section{QER Performance Upper Bound}\label{sec:Dual Bounds}

In the preceding sections, we imposed constraints on the recovery operations to provide structure and aid computation.  While the resulting channel-adapted recoveries out perform the generic QEC recovery operation in all of the examples, the constraints essentially guarantee sub-optimality.  For the five qubit code (where computation of the optimal QER operation is practical), we observe that the proposed algorithms (EigQER, BlockEigQER, and OrderQER) closely approximate the optimal performance.  This anecdotal evidence, however, is hardly sufficient to justify the bold description in the  title of `near-optimal' channel-adapted QER.  In this section, we more fully justify the near-optimal label by deriving channel-adapted performance bounds.  We accomplish this by using the Lagrange dual function.

Every optimization problem has an associated dual problem\cite{BoyVan:B04}.  Derived from the objective function and constraints of the original optimization problem (known as the \emph{primal} problem), the dual problem optimizes over a set of dual variables often subject to a set of dual constraints.  The dual problem has several useful properties.  First of all, the dual problem is always convex.  In many cases, calculation of the dual function is a useful method for constructing optimization algorithms.  Most important for our purposes, the dual function provides a bound for the value of the primal function.  We define a \emph{dual feasible point} as any set of dual variables satisfying the dual constraint.  The dual function value for any dual feasible point is less than or equal to the primal function at any primal feasible point.  (We have implicitly assumed the primal function to be a minimization problem, which is the canonical form.)

The dual function for channel-adapted recovery was derived in \cite{KosLid:06}; we will re-derive it here in a notation more convenient for our purposes.

The primal problem as given in \cite{FleShoWin:07} can be stated succinctly as
\begin{equation}\label{eq:primal problem}
  \min_X -\tr X\CC,\textrm{ such that }X\geq0\textrm{ and }\tr_{\HH_S}X=I.
\end{equation}
The negative sign on the $\tr X\CC$ terms casts the primal problem as a minimization, which is the canonical form.
The Lagrangian is given by
\begin{equation}\label{eq:Lagrangian}
  L(X,Y,Z)=-\tr X\CC +\tr Y(\tr_{\HH_S}X-I)-\tr ZX,
\end{equation}
where $Y$ and $Z\geq 0$ are operators that serve as the lagrange multipliers for the equality and generalized inequality constraints, respectively.    The dual function is the (unconstrained) infimum  over $X$ of the Lagrangian:
\begin{eqnarray}\label{eq:dual_infimum}
  g(Y,Z)&=&\inf_X L(X,Y,Z)\\\label{eq:dual function}
  &=&\inf_{X} -\tr{X(\CC+Z-I\otimes Y)} - \tr Y,
\end{eqnarray}
where we have used the fact that $\tr (Y\tr_{\HH_S}X)=\tr (I\otimes Y) X$.  Since $X$ is unconstrained, note that $g(Y,Z)=-\infty$ unless $Z=I\otimes Y-\CC$ in which case the dual function becomes $g(Y,Z)=-\tr Y$.  $Y$ and $Z\geq 0$ are the dual variables, but we see that the dual function depends only on $Y$.  We can therefore remove $Z$ from the function as long as we remember the constraint implied by $Z=I\otimes Y-\CC$.  Since $Z$ is constrained to be positive semidefinite, this can be satisfied as long as $I\otimes Y-\CC\geq 0$.

We now have the bounding relation $-\tr X\CC\geq \tr -Y$ for all $X$ and $Y$ that are primal and dual feasible points, respectively.  If we now reverse the signs so that we have a more natural fidelity maximization, we write
\begin{eqnarray}\label{eq:primal dual inequality}
  \bar{F}_e(E_,\R\circ\E)=\tr X_\R \CC\leq \tr Y,
\end{eqnarray}
where $\R$ is CPTP and $I\otimes Y-\CC\geq 0$.  To find the best bounding point $Y$, we solve the dual optimization problem
\begin{equation}\label{eq:dual problem}
  \min_Y \tr Y,\textrm{ such that }I\otimes Y-\CC\geq 0.
\end{equation}
Notice that the constraint implies that $Y=Y^\dagger$.  Note also that $Y\in\LL(\HH_C^*)$.

We will use the bounding property (\ref{eq:primal dual inequality}) of the dual function.  Given any dual feasible point $Y\in \LL(\HH_C^*)$, we know that $\tr Y$ upper bounds $\bar{F}_e(E,\R\circ\E)$ for all $\R$; $Y$ is thus a certificate of convergence for a recovery operation.

To provide a good performance bound, it is desirable to find a dual feasible point with a small dual function value.  Indeed, the best such bound is the solution to (\ref{eq:dual problem}), that is to find the dual feasible point with the smallest trace.  However, finding the optimal $Y$ is the equivalent of solving for the optimal recovery due to the strong duality of the SDP.  As this suffers the same computational burden as computing the optimal recovery, we require an alternate method for generating useful dual feasible points.  We will establish methods to convert the sub-optimal recovery operations of the preceding sections into dual feasible points.

We need to determine a good dual feasible point beginning with one of the sub-optimal recoveries computed by the EigQER, BlockEigQER, or OrderQER algorithms.
We utilize the structure of the sub-optimal recovery operations to generate a dual feasible point.  We present two methods that exploit the projective syndrome measurement to achieve performance bounds.  The first bound is motivated by the proof of Theorem \ref{thm:PauliChannel} in the appendix, where the optimal dual feasible point is constructed for Pauli group errors.  Beginning with this construction and the recovery generated by EigQER, we use the Ger\v{s}gorin disc theorem to generate a dual feasible point.  The resulting dual function we denote the Ger\v{s}gorin dual bound.  The second construction iteratively generates dual feasible points given an initial infeasible point.  While it is more computationally burdensome, it generates tighter bounds for the considered examples.  We begin with a trial dual variable that may or may not be feasible and iteratively extend this point until it is feasible.  We call this construction the iterative dual bound.  We present several methods for providing an initial trial point.

Discussion of both bounding methods is facilitated by choosing an appropriate basis for $\HH_S\otimes \HH_C^*$.  Both methods begin with a recovery operation generated by one of the structured sub-optimal methods.  As they all begin with a projective measurement, the recovery provides a partition of $\HH_C$ into subspaces $\SSS_q$ of dimension $d_q$ described by projection operators $\{P_q\}\in\LL(\HH_C)$.  We are interested in a basis $\{\ket{v_i}\}_{i=1}^{2^{n+k}}$ where the first block of $d_Sd_0$ basis vectors span $I\otimes\SSS_0^*$ and the $q^{th}$ block spans $I\otimes\SSS_q^*$.  Let us define
\begin{equation}
(\CC)_{qq'}\equiv I\otimes \Ol{P_q}\CC I\otimes \Ol{P_{q'}}
\end{equation}
as we did in (\ref{eq:reduced SDP}) and then write
\begin{equation}
  \CC=\begin{bmatrix}
    (\CC)_{00} & \cdots & (\CC)_{0q} & \cdots\\
    \vdots & \ddots &  \vdots & \\
    (\CC)_{q0} & \cdots& (\CC)_{qq} &\\
    \vdots & & & \ddots
  \end{bmatrix}
\end{equation}
in our defined basis.
This block structure delineates the relationship of the data operator $\CC$ on each of the subspaces $\SSS_q$ which will be useful when discussing dual feasible points.

\subsection{Ger\v{s}gorin Dual Bound}

The first method for constructing dual feasible points imposes a convenient structure on $Y$.  In the case of Pauli group errors considered in \cite{Fle:07}, the optimal dual feasible point has the form
\begin{equation}\label{eq:dual feasible form}
  Y=\sum_qw_q \Ol{P_q},
\end{equation}
where $w_q$ are a set of weights corresponding to the probability of the most likely error resulting in the $q^{th}$ syndrome measurement.  The form of (\ref{eq:dual feasible form}) is appealing due its simplicity, especially for the EigQER recovery operation where the rank $d_q$ of the $P_q$ is constrained to be $\leq d_S$.  While we cannot necessarily generate the optimal dual feasible point in this form for non-Pauli errors, we can use similar methods to generate a reasonable performance bound.

Before we state the Ger\v{s}gorin dual bound, we take a second look at the optimal dual point for Pauli errors.  For an $[n,k]$ stabilizer code, recall that $\HH_C$ is partitioned into $2^{n-k}$ syndrome subspaces $\SSS_q$ and we establish a basis $\{\ketsub{m}{q}\}$ for each subspace.  We also determined that $\kett{U_{Cq}^\dagger A_p}$ is an eigenvector of $\CC$.  Note that $\{\kett{U_{Cq}^\dagger A_p}\}_{p=0}^{2^{2k}-1}$ span the space $I\otimes\Ol{\SSS_q}$.

If we write out the operator $(\CC)_{qq}$ in this basis, we have
\begin{eqnarray}
  (\CC)_{qq}&=&\begin{bmatrix}
  a_{0q} &&\\
  & \ddots &\\
  & & a_{(2^{2k}-1)q}
  \end{bmatrix}
\end{eqnarray}
which is diagonal because $\{\ketsub{m}{q}\}$ are eigenvectors of $\CC$.  This also implies that all of the off-diagonal blocks $(\CC)_{qq'}$ where $q\neq q'$ are also 0.   We can now see that $Y=\sum_q \tilde{a}_q\Ol{P_q}$ where $\tilde{a}_q=\max_p |a_{pq}|$ is a dual feasible point since
\begin{equation}
  I\otimes Y^\star = \begin{bmatrix} \tilde{a}_0 I & 0 & \cdots & 0\\ 0 &\tilde{a}_1I & \cdots& 0\\ \vdots & \vdots  & \ddots & \vdots\\
  0 & 0 & \cdots & \tilde{a}_{2^{n-k}-1}I\end{bmatrix}\\
\end{equation}
is diagonal in the chosen basis.

We return now to the general case.  Unlike in the case of a Pauli error channel and a stabilizer code, we cannot guarantee that $\CC$ will be either diagonal or block diagonal in this basis.  However, if our sub-optimal recovery $\R$ is generated from the EigQER algorithm, then the subspaces $\SSS_q$ are selected based on the eigenvectors of $\CC$ and we can expect $\CC$ to be approximately block diagonal when we partition according to the subspaces $I\otimes \SSS_q^*$.  We say that $\CC$ is approximately block diagonal in this basis if $\|(\CC)_{qq}\|\gg\|(\CC)_{qq'}\|$ for $q\neq q'$.

To generate a dual feasible point of the form $Y=\sum_q w_q \Ol{P_q}$, we need to choose $w_q$ so that $I\otimes Y-\CC\geq 0$.  If $\CC$ were exactly block diagonal in this basis, we could accomplish this by setting $w_q=\lambda_{\max}((\CC)_{qq})$.  Since the block terms off the diagonal are not strictly 0, we must account for their contributions in the location of the eigenvalues of $\CC$.

We will make use of a linear algebra theorem known as the Ger\v{s}gorin disc theorem.  This theorem provides bounds on the location in the complex plane of the eigenvalues of an arbitrary matrix.  As will be evident, the theorem is most valuable when the matrix is dominated by its diagonal entries.  We state the theorem as it is given in \cite{HorJoh:B85} $\S$ 6.1:

\begin{Gersgorin}
  \label{thm:Gersgorin}
  Let $A=[a_{ij}]\in \mathbb{C}^{n\times n}$, and let
  \begin{equation}
    R_i'(A)\equiv\sum_{j=1, j\neq i}^n |a_{ij}|,\hspace{20pt}1\leq i\leq n
  \end{equation}
denote the \emph{deleted absolute row sums} of $A$.  Then all the eigenvalues of $A$ are located in the union of $n$ discs
\begin{equation}
  \bigcup_{i=1}^n \{ z\in\mathbb{C}:|z-a_{ii}|\leq R_i'(A)\}\equiv G(A).
\end{equation}
Furthermore, if a union of $k$ of these $n$ discs forms a connected region that is disjoint from all the remaining $n-k$ discs, then there are precisely $k$ eigenvalues of $A$ in this region.
\end{Gersgorin}

Theorem \ref{thm:Gersgorin} is particularly useful for proving the positivity of a matrix.  The $R_i'(A)$ are the radii of discs centered at the diagonal entries $a_{ii}$ and the eigenvalues are constrained to lie within the union of these discs.  If $A$ is a Hermitian matrix, then we can be certain it is positive semidefinite if $a_{ii}\geq R_i'(A)$ for all $i$ as all of the eigenvalues would be constrained to lie to the right of the origin (or on the origin) on the real line.

We can apply Theorem \ref{thm:Gersgorin} to generating a dual feasible point structured as (\ref{eq:dual feasible form}).  In this case we use the weights $w_q$ to ensure that the diagonal entries of $I\otimes Y-\CC$ are greater than the deleted absolute row sums.  Let $c_{ij}$ denote the matrix elements of $\CC$ in our defined basis and let the basis vector $\ket{v_i}$ lie in the subspace $\SSS_q$.  We then the have the $i^{th}$ diagonal element $[I\otimes Y-\CC]_{ii}=w_q-c_{ii}$ and the $i^{th}$ deleted absolute row sum is $\sum_{i\neq j} |c_{ij}|$.  We can assure non-negativity if
\begin{equation}
  w_q\geq \sum_j |c_{ij}|,\textrm{ for all }i\textrm{ such that }\ket{v_i}\in \SSS_q.
\end{equation}
Thus, we can guarantee a dual feasible point if $w_q$ is set to be the maximum absolute row sum for all rows $i$ such that $\ket{v_i}\in\SSS_q$.
We may express $w_q$ concisely in terms of the induced $\infty$-norm(\cite{HorJoh:B85} $\S$ 5.6.5), denoted $\|\cdot\|_\infty$:
\begin{eqnarray}
  w_q&=&\left\|\begin{bmatrix} (\CC)_{q0} & \cdots & (\CC)_{qq} & \cdots \end{bmatrix}\right\|_\infty\\
  &=&\|I\otimes\Ol{P_q}\CC\|_\infty.
\end{eqnarray}

\begin{figure}
    \includegraphics[width=\columnwidth]{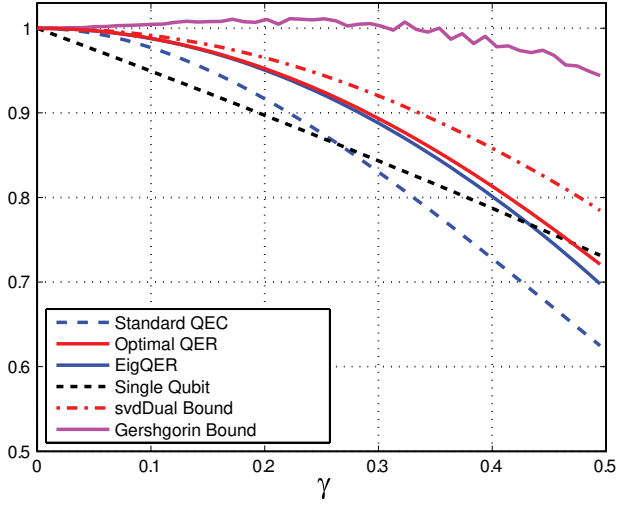}
  \caption[Ger\v{s}gorin and SVD dual bound for the amplitude damping channel and the 5 qubit stabilizer code.]{Ger\v{s}gorin and SVD dual bound for the amplitude damping channel and the 5 qubit stabilizer code.  The Ger\v{s}ogrin bound is clearly not very useful as in some cases it is greater than 1.  The SVD dual bound clearly tracks the optimal performance, although the departure from optimal of the bound exceeds the EigQER recovery.}\label{fig:EigDUAL 5}
\end{figure}

The Ger\v{s}gorin disc theorem is a computationally simple way to guarantee construction of a dual feasible point given a partition of $\HH_C$ into subspaces $\{\SSS_q\}$.  Unfortunately, the induced infinity norm does not provide a particularly useful performance bound as can be seen in Figure
\ref{fig:EigDUAL 5}.  When we compare to the optimal recovery performance for the five qubit code and the amplitude damping channel, we see that the dual bound is far from tight.  In fact, for many values of $\gamma$, the bound is greater than 1, which is truly useless for upper bounding fidelities.  While we have generated a dual point $Y$ that is guaranteed to be feasible, such a guarantee imposes too strict a cost to have a useful bounding property.

The Ger\v{s}gorin dual bound provides useful insight for a tighter dual construction.  If we replace the induced infinity norm with the induced 2-norm, we generate a dual point that is often dual feasible.  That is, choose
\begin{eqnarray}
  w_q&=&\|I\otimes\Ol{P_q}\CC\|_2\\
  &=& \max_{\kett{x}} \braa{x}I\otimes\Ol{P_q}\CC\kett{x}\\
  \label{eq:svd 2norm}
  &=& \sigma_{\max}(I\otimes\Ol{P_q}\CC),
\end{eqnarray}
where $\sigma_{\max}(\cdot)$ in (\ref{eq:svd 2norm}) indicates the maximum singular value and is the computational method for the induced 2-norm.  We will refer to this construction as the SVD (for singular value decomposition) dual point.  The $Y$ generated in this way is not guaranteed to be dual feasible as was the case with the $\infty$-norm, but has proven to be dual feasible in all of the examples that we have tried.  If for some circumstance the SVD dual point is not feasible, it can be iteratively adjusted to become dual feasible in a manner we present in the following section.

\subsection{Iterative Dual Bound}\label{sec:IterativeDual}

We now present an iterative procedure to generate a dual feasible point given an initial dual point $Y^{(0)}$ that is presumably not dual feasible.  After presenting the algorithm, we will discuss choices for the initial dual point.

At the $k^{th}$ iteration, we update the dual point to produce $Y^{(k)}$ until we achieve feasibility.  For convenience we will define
\begin{equation}
Z^{(k)} \equiv I\otimes Y^{(k)}-\CC.
\end{equation}
Let $x$ and $\kett{x}$ be the smallest eigenvalue and associated eigenvector of $Z^{(k)}$.  If $x\geq 0$, we may stop, as $Y^{(k)}$ is already dual feasible.  If $x\leq 0$, we wish to update $Y^{(k)}$ a small amount to ensure that $\braa{x}Z^{(k+1)}\kett{x}\geq 0$.  Essentially, we are replacing a negative eigenvalue with a 0 eigenvalue.  Given no constraints on the update, we could accomplish this as $Z^{(k+1)}=Z^{(k)}+x\kett{x}\braa{x}$ but we must instead update $Y^{(k)}$ with the tensor product structure implicit.

We determine the properly constrained update by means of the Schmidt decomposition of the eigenvector:
\begin{equation}
\kett{x}=\sum_i \lambda_i\ketsub{\hat{x}_i}{\HH_S}\ketsub{\tilde{x}_i}{\HH_C^*}.
\end{equation}
As we can only perturb $Z^{(k)}$ in the $\HH_C^*$ slot, we choose the smallest perturbation guaranteed to achieve $\braa{x}Z^{(k+1)}\kett{x}\geq 0$.
Let
\begin{equation}
Y^{(k+1)}=Y^{(k)}+\frac{|x|}{|\lambda_1|^2}\ket{\tilde{x}_1}\bra{\tilde{x}_1}.
\end{equation}
Then
\begin{eqnarray}
  \braa{x}Z^{(k+1)}\kett{x}&=& x+\frac{|x|}{|\lambda_1|^2}\braa{x}(I\otimes \ket{\tilde{x}_1}\bra{\tilde{x}_1})\kett{x}\\
  &=& x+\frac{|x|}{|\lambda_1|^2}|\lambda_1|^2\\
  &=& 0,
\end{eqnarray}
since $x<0$.
While we have not yet guaranteed that $Z^{(k+1)}\geq 0$, $\kett{x}$ is no longer associated with a negative eigenvalue.  By repeatedly perturbing $Y^{(k)}$ in this manner, we iteratively approach a dual feasible point while adding as little as possible to the dual function value $\tr Y^{(k)}$.

As a final point, we demonstrate that the iterative procedure will converge to a dual feasible point.  Let's consider the effect of the $k^{th}$ iteration on the space orthogonal to $\kett{x}$.  Let $\kett{y}\in\HH_S\otimes\HH_C^*$ be orthogonal to $\kett{x}$.  Then, for $Z^{(k+1)}$ we see that
\begin{eqnarray}
  \braa{y}Z^{(k+1)}\kett{y}=\braa{y}Z^{(k)}\kett{y}+\frac{|x|}{|\lambda_1|^2}\braa{y}(I\otimes \ket{\tilde{x}_1}\bra{\tilde{x}_1})\kett{y}.
\end{eqnarray}
But since $I\otimes \ket{\tilde{x}_1}\bra{\tilde{x}_1}\geq 0$ we see that
\begin{equation}
  \braa{y}Z^{(k+1)}\kett{y}\geq\braa{y}Z^{(k)}\kett{y}
\end{equation}
for all $\kett{y}\in\HH_S\otimes\HH_C^*$.  We see that the update to $Y^{(k)}$ moved one negative eigenvalue to 0 while no new negative eigenvalues can be created.  Thus the procedure will eventually converge to a dual feasible point.

\subsection{Initial dual points}

Having established a procedure to generate a dual feasible point given an arbitrary intial point $Y^{(0)}$, we now present initialization options.  While we can start with any Hermitian operator in $\LL(\HH_C^*)$ including $0$, we do not recommend such an unstructured choice as each iteration is imperfect.  Each iteration adds $|x|/|\lambda_1|^2$ to the dual function value.  If $|\lambda_1|$ is not close to 1, the iteration is not efficient.  We will use more educated initializations to begin closer to feasibility, thus minimizing the number of iterations and improving the bounding properties of the resulting dual feasible point.

We have already presented one method for initialization with the SVD dual point.  In most cases we've seen, this point is already feasible and in fact is a relatively loose bound.  Its advantage lies in its easy computation, but other choices provide better bounding properties.  We would prefer an initial $\Yi$ such that $\Zi$ is non-positive with eigenvalues very close to 0.  If this is the case, we will require only small perturbations (and thus a small dual function value) to achieve a positive semidefinite $Z^{(k)}$.

Consider an initial $Y^{(0)}$ of the form given in (\ref{eq:dual feasible form}).  We choose an initial $\Yi$ in the same way that was used in the proof of Theorem \ref{thm:PauliChannel}:
\begin{equation}
  w_q=\lambda_{\max}((\CC)_{qq}).
\end{equation}
This is very simple to calculate, though it will not generally be dual feasible.  This is the logical choice when we begin with the EigQER recovery, as the only useful information we have is the projective syndrome measurement.  This initialization often iterates to a better bound than the SVD dual point and requires no further information than the partition $\{\SSS_q\}$ provided by any of the structured QER methods.  It has one drawback, however, in that $\Zi$ almost certainly has eigenvalues much greater than 0.  For the $\ket{v_i}$ associated with the largest eigenvalue of $(\CC)_{qq}$, $\bra{v_i}\Zi\ket{v_i}=0$.  However, unless $(\CC)_{qq}$ has only one distinct eigenvalue there will be vectors $\kett{x}\in\SSS_q$ such that $\braa{x}\Zi\kett{x}\geq0$, and perhaps quite large, relatively.  Such vectors indicate portions of the Hilbert space where $\Yi$ is already greater than the optimal dual feasible point.  While this likely cannot be avoided in the iterations, it seems wasteful to begin at such a point if not necessary.

We have an alternative choice for $\Yi$ arising from the block SDP QER algorithms of Sec.~\ref{sec:BlockSDP}.  These algorithms already provide information useful for generating a dual feasible point.  When solving the SDP on a subspace $\SSS_q$ one can simultaneously generate the optimal dual function value $Y_q^\star\in\LL(\SSS_q^*)$.   Given such optimal subspace dual points, define the block diagonal operator
\begin{equation}
  \Yi=\begin{bmatrix}
    Y_0^\star &&&\\
    & \ddots &&\\
    && Y_q^\star &\\
    &&&\ddots
  \end{bmatrix}
\end{equation}
as the initial point.  We know that $I\otimes Y_q^\star -(\CC)_{qq}\geq 0$, so there will be $\kett{x}$ for which $\braa{x}\Zi\kett{x}\geq0$.  However, since $Y_q^\star$ is optimal within $\LL(\SSS_q^*)$, we know that we are not being overly wasteful with the initialization.

\subsection{Iterated block dual}

Let's consider the computational burden of the iterated dual bound.  At each iteration we must compute the smallest eigenvalue and associated eigenvector of $Z^{(k)}$, a $2^{n+k}\times 2^{n+k}$ Hermitian matrix.  (We can accomplish this by looking for the largest eigenvalue of $\eta I-Z^{(k)}$ where $\eta\geq1$ is an arbitrary offset to ensure positivity.)  This must be repeated at most $2^{n+k}$ times to ensure dual feasibility, though there may be significantly fewer iterations if the $Z^{(0)}$ is nearly positive semidefinite already.  As mentioned in Sec.~\ref{sec:EigQER}, this can be accomplished in $\mathcal{O}(2^{2(n+k)})$ flops by the power method.  This is very costly if we must repeat the iteration many times.

The block diagonal structure of the initial points suggests a slightly modified alternative procedure with some computational advantage. Consider the optimal dual points $Y_i$ and $Y_j$ in $\LL(\SSS_i^*)$ and $\LL(\SSS_j^*)$.  We can use the same iterative procedure as before to compute a dual feasible $Y_{ij}\in\LL(\SSS_i^*\oplus\SSS_j^*)$ requiring only $\mathcal{O}(2^{2k}(d_i+d_j)^2)$ flops per iteration with a maximum of  $2^{k}(d_i+d_j)$ iterations.  We can generate a dual feasible point on the whole space $\LL(\HH_C^*)$ by successively combining subspace blocks.  Eventually we will have to iterate over the full space, but we will have done most of the work in the smaller blocks, and the full $2^{n+k}\times 2^{n+k}$ eigen decomposition will require few iterations.

In the examples we have processed, the iterated block dual procedure created nearly identical bounds (often within $10^{-5}$ of each other and never more than $10^{-4}$) as the original algorithm.  The computational burden is reduced by approximately $20\%$.

\subsection{Examples}

\begin{figure}
    \includegraphics[width=\columnwidth]{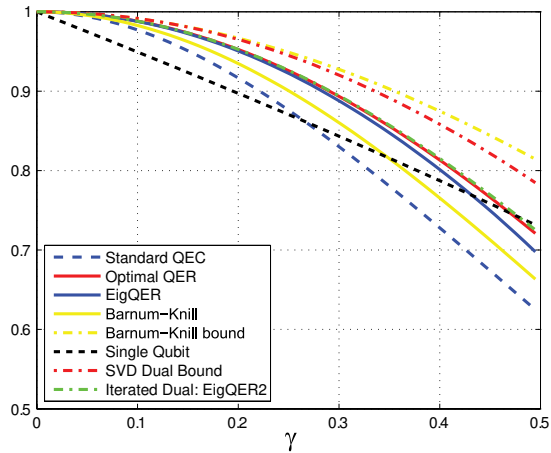}
  \caption[Dual bound comparison for the amplitude damping channel and the five qubit code.]{Dual bound comparison for the amplitude damping channel and the five qubit code.  The iterated dual initialized with the Block EigQER algorithm with $M=2$ is essentially indistinguishable from the optimal recovery performance, thus producing a very tight bound.  Included for comparison are the EigQER performance, the SVD dual bound, and both a channel-adapted recovery and associated bound derived by Barnum and Knill in \cite{BarKni:02}.}\label{fig:AmpDamp5_dual}
\end{figure}

We provide several examples to demonstrate the utility of the iterated dual bound.  At the same time, we we illustrate the near optimality of the structured QER algorithms.  In Fig.~\ref{fig:AmpDamp5_dual}, we show several bounds for channel-adapted QER for the amplitude damping channel and the five qubit code.  In this case, we know the optimal performance and can see that the iterated dual bound,  beginning with the BlockEigQER with $M=2$, is quite tight.  This is in contrast to the SVD dual bound, which was also shown in Fig.~\ref{fig:EigDUAL 5}.  We have included in Fig~\ref{fig:AmpDamp5_dual} the numerical channel-adapted recovery and performance bound from \cite{BarKni:02}.  We see that this bound is looser than even the SVD dual bound for this example.

\begin{figure}
    \includegraphics[width=\columnwidth]{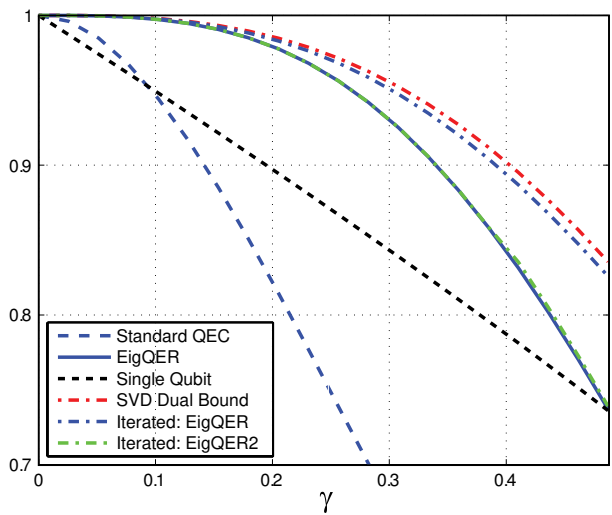}
  \caption[Dual bound comparison for the amplitude damping channel and the nine qubit Shor code.]{Dual bound comparison for the amplitude damping channel and the nine qubit Steane code.  The iterated dual bound initialized with the BlockEigQER recovery with $M=2$ produces a bound that is tight to the EigQER recovery operation.  This demonstrates that the EigQER recovery operation is essentially optimal in this case.  Notice that the iterated bound initialized with the EigQER recovery operation does not generate a tight bound.}\label{fig:AmpDamp9_dual}
\end{figure}

Figure \ref{fig:AmpDamp9_dual} shows several dual bounds for the amplitude damping channel and the nine qubit Shor code.  While we cannot compute the optimum directly, we see that the EigQER performance curve and the iterated bound derived from BlockEigQER with $M=2$ are essentially equivalent.  We can conclude that EigQER operation is essentially optimal in this case.  While not shown, iterations for BlockEigQER with $M=4$ and $M=8$ achieved essentially the same bound.  Note that neither the SVD dual bound nor the iterated bound beginning with the EigQER recovery operation are tight, illustrating the importance of a good initialization for the dual iterations.

\begin{figure}
    \includegraphics[width=\columnwidth]{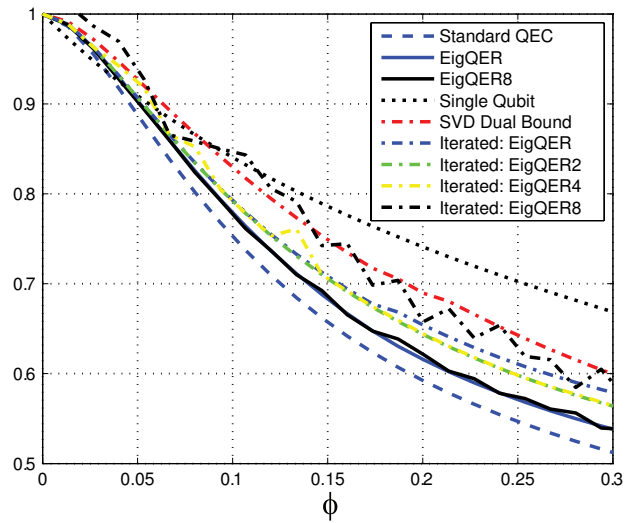}
\caption[Dual bound comparison for the pure state rotation channel with $\theta=5\pi/12$ and the seven qubit Steane code.]{Dual bound comparison for the pure state rotation channel with $\theta=5\pi/12$ and the seven qubit Steane code.  Note that the iterated bounds are generally, though not universally, better than the SVD dual bound.  We also see that the shorter block lengths for the BlockEigQER algorithm generally produce a tighter bound, despite slightly poorer recovery performance.}\label{fig:PureState7_dual}
\end{figure}

Our final example is the pure state rotation channel with $\theta=5\pi/12$ and the seven qubit Steane code.  In Fig.~\ref{fig:PureState7_dual}, we can distinguish between several initialization methods for the dual iterative bound.  We see that none of the recovery operations approach the bound performance for large $\phi$, though the performance is relatively tight as the noise level drops ($\phi\rightarrow 0)$.  Notice that in general the iterative bounds are better than the SVD dual bound, however there are points, especially for the BlockEigQER algorithm with $M=8$, where the iterated bound is poor.  It is interesting to note that the longer block lengths (larger $M$) usually generate better recovery performance (which can be seen with slight improvement even in this case) yet often produce poorer bounds.  Anecdotal experience suggests that the best iterative starting point is the BlockEigQER recovery operation with $M=2$.

Finally, we should point out the gap for large $\phi$ between the recovery performance and the dual bounds.  Absent a better recovery operation or a smaller performance bound, we have no way to know whether the bound or the recovery is further removed from the optimal.  However, this region is below the baseline performance for a single unencoded qubit, and thus is not of serious concern.

\section{Conclusion and Future Work}

Adapting a quantum recovery operation to a physical channel can significantly improve the effectiveness of a quantum channel.  In this way, quantum error correction can be made more efficient, which should aid in scaling physical implementations to a larger number of qubits.  While the optimal recovery (in terms of average entanglement fidelity) may be calculated via convex optimization of a semidefinite program, we have derived a class of near-optimal algorithms that are less computationally intensive.  Furthermore, these algorithms yield recovery operations of a particular form: they implement a projective error syndrome measurement followed by a syndrome recovery operation.  This structure may prove easier to implement physically and provides intuition into the mechanism for channel-adaptation.

Despite the reduction in computation from the SDP, even these algorithms grow exponentially in the length of (\emph{i.e.}~the number of qubits in) the code.  For this reason, the next step toward practical application of channel-adapted quantum error correction must include analytical tools to supplement these numerical techniques.  Furthermore, to apply channel-adapted methods to fault-tolerant quantum computing, we must show how errors propagate from block to block.  These two open questions are likely closely linked.  Despite these obstacles, the added efficiency of channel-adapted recovery suggests significant value for practical efforts in quantum error correction.

\begin{acknowledgments}
This paper is based on a thesis submitted in partial fulfillment of the requirements for the degree of Doctor of Philosophy in the Department of Electrical Engineering and Computer Science at the Massachusetts Institute of Technology in June, 2007.

A.S.F. would like to thank the Department of the Air Force, who sponsored this work under AF Contract \#FA8721-05-C-0002.   All authors thank the National Science Foundation for support through grant CCF-0431787.  Opinions, interpretations, recommendations and conclusions are those of the authors and are not necessarily endorsed by the United States Government.
\end{acknowledgments}

\bibliography{quantum}

\section{Appendix}

The discussion of the Ger\v{s}gorin and SVD dual bounds make use of a structured dual feasible point.  This is motivated by the optimal dual feasible point for a stabilizer code and a Pauli error channel.  Construction of this optimal dual feasible point proves the intuitive structure of the optimal recovery operation for Pauli error channels.  This theorem was proven in \cite{Fle:07} and will be restated here for reference.

We can construct the optimal recovery operation for a stabilizer code when the channel $\E'$ is characterized by Pauli group errors and the input ensemble is the completely mixed state.  That is, $E$ is given by $\rho=I/d_S$ with $p=1$ and the channel can be represented by Kraus operators $\{E_i\}$ where each $E_i$ is a scaled element of the Pauli group.  (Notice that this does not require every set of Kraus operators that characterize $\E'$ to be scaled elements of the Pauli group, since unitary combinations of Pauli group elements do not necessarily belong to the Pauli group.)

To state the optimal recovery, we carefully define the syndrome measurement subspaces and the Pauli group operators that connect the subspaces.  We must do this in a way to consistently describe the normalizer operations of the code.  Consider an $[n,k]$ stabilizer code with generators $\langle g_1,\ldots,g_{n-k}\rangle$ and logical $\bar{Z}$ operators $\bar{Z}_1,\ldots\bar{Z}_k$ such that $\{g_1,\ldots,g_{n-k},\bar{Z}_1,\ldots\bar{Z}_k\}$ form an independent and commuting set.  Define logical $\bar{X}$ operators such that $[\bar{X}_i,g_j]=[\bar{X}_i,\bar{X}_j]=0$ $\forall$ $i,j$, $[\bar{X}_i,\bar{Z}_j]=0$ for $i\neq j$ and $\{\bar{X}_i,\bar{Z}_i\}=0$.

The syndrome subspaces correspond to the intersection of the $\pm 1$ eigenspaces of each generator.  Accordingly, we label each space $\SSS_q$ where $q=0,1,\ldots,2^{n-k}-1$, where $\SSS_0$ corresponds to the code subspace.  Let $P_q$ be the projection operator onto $\SSS_q$.  Let $\{\ketsub{i_1 i_2\cdots i_k}{q}\}$ form a basis for $\SSS_q$ such that
\begin{equation}
  \bar{Z}_1\bar{Z}_2\cdots\bar{Z}_k\ketsub{i_1 i_2\cdots i_k}{q}=(-1)^{i_1}(-1)^{i_2}\cdots(-1)^{i_k}\ketsub{i_1 i_2\cdots i_k}{q},
\end{equation}
where $i_j\in\{0,1\}$.  In this way, we have a standardized basis for each syndrome subspace which can also be written as $\{\ketsub{m}{q}\}$, $m=0,\ldots,2^k-1$.

Let us recall the effect of a unitary operator on a stabilizer state.  If $\ket{\psi}$ is stabilized by $\langle g_1,\ldots,g_{n-k}\rangle$, then $U\ket{\psi}$ is stabilized by $\langle Ug_1U^\dagger,\ldots,Ug_{n-k}U^\dagger\rangle$.  What happens if $U\in G_n$, the Pauli group on $n$ qubits?  In that case, since U either commutes or anti-commutes with each stabilizer, $U\ket{\psi}$ is stabilized by $\langle \pm g_1,\ldots,\pm g_{n-k}\rangle$ where the sign of each generator $g_i$ is determined by whether it commutes or anti-commutes with $U$.  Thus, a Pauli group operator acting on a state in the code subspace $\SSS_0$ will transform the state into one of the subspaces $\SSS_q$.

We have established that the Pauli group errors always rotate the code space onto one of the stabilizer subspaces, but this is not yet sufficient to determine the proper recovery.  Given that the system has be transformed to subspace $\SSS_q$, we must still characterize the error by what happened within the subspace.  That is to say, the error consists of a rotation to a syndrome subspace and a normalizer operation within that subspace.

Let us characterize these operations using the bases $\{\ketsub{m}{q}\}$.  Define $W_{qq'}\equiv\sum_m\ketsub{m}{q'}\brasub{m}{q}$ as the operator which transforms $\SSS_q\mapsto \SSS_{q'}$ while maintaining the ordering of the basis.  Define the encoding isometry $U_C\equiv\sum_m \ketsub{n}{0}\brasub{n}{S}$ where $\ketsub{n}{S}\in\HH_S$, the source space.  Further define $U_{cq}\equiv W_qU_C$, the isometry that encodes the $q^{th}$ syndrome subspace.  We will define the $4^k$ code normalizer operators as
\begin{equation}
  A_p\equiv \bar{X}_1^{i_1}\bar{X}_2^{i_2}\cdots\bar{X}_k^{i_k}
  \bar{Z}_1^{j_1}\bar{Z}_2^{j_2}\cdots\bar{Z}_k^{j_k}
\end{equation}
where $p$ is given in binary as $i_1i_2\cdots i_kj_1j_2\cdots j_k$.  Notice that if a similarly defined $A_p^S$ is an element of the Pauli group $\mathcal{G}_k\in\LL(\HH_S)$ with generators $\langle X_1^S,\ldots,X_k^S,Z_1^S,\ldots,Z_k^S\rangle$, we can conclude $A_pU_C=U_CA_p^S$.

The preceding definitions were chosen to illustrate the following facts.  First, we can see by the definitions that $[W_{qq'},A_p]=0$.  That is, $W_{qq'}$ characterizes a standard rotation from one syndrome subspace to another, and $A_p$ characterizes a normalizer operation within the subspace.  These have been defined so that they can occur in either order.  Second, let $\E'$ be a quantum channel represented by operator elements that are scaled members of the Pauli group $\mathcal{G}_n$.  Then the composite channel $\E$ which includes the encoding isometry $U_C$ can be represented by operator elements of the form
\begin{equation}\label{eq:Pauli_channel}
  \{E_{pq}=a_{pq}A_pW_qU_C=a_{pq}A_pU_{Cq}\},
\end{equation}
where the CPTP constraint requires $\sum_{pq}|a_{pq}|^2=1$.

We can understand the amplitudes $a_{pq}$ by noting that with probability $|a_{pq}|^2$, the channel $\E$ transforms the original state to $\SSS_q$ and applies the normalizer operation $A_p$.  To channel-adaptively recover, we project onto the stabilizer subspaces $\{\SSS_q\}$ and determine the most likely normalizer operation for each syndrome subspace $\SSS_q$.  Let $p_q=\arg\max_p|a_{pq}|^2$, and let $\tilde{a}_q\equiv a_{p_qq}$.  With these definitions in place, we can state the following theorem:
\begin{Pauli_Channel}\label{thm:PauliChannel}
Let $\E$ be a channel in the form of (\ref{eq:Pauli_channel}), i.e.~a stabilizer encoding and a channel with Pauli group error operators.  For a source in the completely mixed state $\rho=I/d_S$ the optimal channel-adapted recovery operation is given by $\R\sim\{U_{Cq}^\dagger A_{p_q}\}$, which is the stabilizer syndrome measurement followed by maximum likelihood  normalizer syndrome correction.
\end{Pauli_Channel}
\begin{proof}
We prove Theorem \ref{thm:PauliChannel} by constructing a dual feasible point $Y$ such that the dual function value $\tr Y$ is equal to the entanglement fidelity $F_e(\rho,\R\circ\E)$.

We begin by calculating $F_e(\rho,\R\circ\E)$.  For later convenience, we will do this in terms of the Choi matrix $\CC$ from (\ref{eq:fidelity max}).
We write the entanglement fidelity in terms of the recovery operator elements $\kett{U_{Cq}^\dagger A_{p_q}}$:
\begin{eqnarray}
  F_e(\rho,\R\circ\E)&=& \tr X_\R \CC\\
  &=&\sum_{q'}\braa{U_{Cq'}^\dagger A_{p_{q'}}}\CC\kett{U_{Cq'}^\dagger A_{p_{q'}}}.\label{eq:proof_ent_fid}
\end{eqnarray}
To evaluate (\ref{eq:proof_ent_fid}), we note that
\begin{eqnarray}
  \label{eq:proof_trace1}
  \braakett{\rho U_{Cq}^\dagger A_{p}}{U_{Cq'}^\dagger A_{p_{q'}}}&=&
  \tr A_pU_{Cq}\rho U_{Cq'}^\dagger A_{p_{q'}}\\
  &=& \tr A_pW_qU_C\rho U_C^\dagger W_{q'}^\dagger A_{p_{q'}}\\\label{eq:proof_trace2}
  &=&\tr A_pW_{q'}^\dagger W_qU_C\rho U_C^\dagger A_{p_{q'}}\\\label{eq:proof_trace3}
  &=&\delta_{qq'}\tr A_pU_C\rho U_C^\dagger A_{p_{q'}}\\\label{eq:proof_trace4}
  &=&\delta_{qq'}\tr A_p^C\rho A_{p_{q'}}^C.
\end{eqnarray}
We have used the commutation relation $[W_{qq'},A_p]=0$ to arrive at (\ref{eq:proof_trace2}) and the facts that $W_{q'}^\dagger W_q=\delta_{qq'}P_0$ and $P_0U_C=U_C$ to conclude (\ref{eq:proof_trace3}).
Since $\rho=I/d_S$ and $\tr A_p^C A_{p_{q'}}^C=\delta_{pp_{q'}}d_S$, we see that $\tr A_p^C\rho A_{p_{q'}}^C=\delta_{pp_{q'}}$.  Thus,
\begin{equation}\label{eq:proof_braakett}
  \braakett{\rho U_{Cq}^\dagger A_{p}}{U_{Cq'}^\dagger A_{p_{q'}}P_{q'}}=
  \delta_{pp_{q'}}\delta_{qq'}.
\end{equation}
Using (\ref{eq:proof_braakett}), it is straightforward to evaluate (\ref{eq:proof_ent_fid}):
\begin{eqnarray}
  F_e(\rho,\R\circ\E)&=&\sum_{pqq'}
  |a_{pq}|^2|\braakett{\rho U_{Cq}^\dagger A_{p}}{U_{Cq'}^\dagger A_{p_{q'}}}|^2\\
  &=& \sum_{pqq'}|a_{pq}|^2\delta_{qq'}\delta_{pp_{q'}}\\
  &=& \sum_{q}|\tilde{a}_q|^2.
\end{eqnarray}

We now propose the dual point $Y=\sum_q |\tilde{a}_q|^2\Ol{P_q}/d_S$.  Since
\begin{eqnarray}
\tr Y &=& \sum_q |\tilde{a}_q|^2\tr{\Ol{P_q}}/d_S\\
&=& \sum_q |\tilde{a}_q|^2\\
&=& F_e(\rho,\R\circ\E),
\end{eqnarray}
we complete the proof by demonstrating that
\begin{equation}
  I\otimes Y-\CC\geq0,
\end{equation}
\emph{i.e.}~$Y$ is a dual feasible point.  We show this by demonstrating that $I\otimes Y$ and $\CC$ have the same eigenvectors, and that the associated eigenvalue is always greater for $I\otimes Y$.

By the same argument used for (\ref{eq:proof_braakett}), we note that
\begin{equation}
  \braakett{\rho U_{Cq}^\dagger A_{p}}{\rho U_{Cq'}^\dagger A_{p'}}
    =\delta_{pp'}\delta_{qq'}/d_S^2.
\end{equation}
This means that $\kett{\rho U_{Cq}^\dagger A_{p}}$ is an eigenvector of $\CC$ with eigenvalue $|a_{pq}|^2/d_S$.  We normalize the eigenvector to unit length and apply it to $I\otimes Y$:
\begin{eqnarray}
  I\otimes Y\kett{\rho U_{Cq}^\dagger A_{p}/d_S}&=&
  \sum_{q'} |\tilde{a}_{q'}|^2\Ol{P_{q'}}/d_S\kett{\rho U_{Cq}^\dagger A_{p}/d_S}\\
  &=& \frac{1}{d_S}\sum_{q'}|\tilde{a}_{q'}|^2\kett{\rho U_{Cq}^\dagger A_{p}P_{q'}/d_S}\\
  &=& \frac{1}{d_S}|\tilde{a}_{q}|^2\kett{\rho U_{Cq}^\dagger A_{p}/d_S}.
\end{eqnarray}
Thus we see that $\kett{\rho U_{Cq}^\dagger A_{p}}$ is an eigenvector of $I\otimes Y$ with eigenvalue $|\tilde{a}_{q}|^2/d_S\geq |a_{pq}|^2/d_S$ $\forall$ $p$.  Thus $I\otimes Y-\CC\geq 0$ and $Y$ is a dual feasible point.
\end{proof}

\end{document}